\documentclass[twocolumn]{aastex631}
\usepackage{xcolor}
\usepackage{amsmath,bigdelim}

\def\at{AT\,2024qfm}

\def\chandra{\emph{CXO}}
\def\nustar{\emph{NuSTAR}}

\begin{document}
\title{ The Most X-ray Luminous Fast Blue Optical Transient at Early Times: AT\,2024qfm and Inefficient Reprocessing of Its High-Energy Emission}

\correspondingauthor{Nayana A.J.}
\email{nayana@berkeley.edu}

\author[0000-0002-8070-5400]{Nayana A. J.}
\affiliation{Department of Astronomy, University of California, Berkeley, CA 94720-3411, USA}
\affiliation{Berkeley Center for Multi-messenger Research on Astrophysical Transients and Outreach (Multi-RAPTOR), University of California, Berkeley, CA 94720-3411, USA}

\author[0000-0003-4768-7586]{Raffaella Margutti}
\affiliation{Department of Astronomy, University of California, Berkeley, CA 94720-3411, USA}
\affiliation{Department of Physics, University of California, 366 Physics North MC 7300,
Berkeley, CA 94720, USA}
\affiliation{Berkeley Center for Multi-messenger Research on Astrophysical Transients and Outreach (Multi-RAPTOR), University of California, Berkeley, CA 94720-3411, USA}

\author[0000-0002-7706-5668]{Ryan Chornock} 
\affiliation{Department of Astronomy, University of California, Berkeley, CA 94720-3411, USA}
\affiliation{Berkeley Center for Multi-messenger Research on Astrophysical Transients and Outreach (Multi-RAPTOR), University of California, Berkeley, CA 94720-3411, USA}

\author[0000-0002-5872-6061]{James F. Steiner}
\affiliation{Center for Astrophysics \textbar{} Harvard \& Smithsonian, 60 Garden Street, Cambridge, MA 02138-1516, USA}

\author[0000-0002-5698-8703]{Erica Hammerstein}
\affiliation{Department of Astronomy, University of California, Berkeley, CA 94720-3411, USA}
\affiliation{Berkeley Center for Multi-messenger Research on Astrophysical Transients and Outreach (Multi-RAPTOR), University of California, Berkeley, CA 94720-3411, USA}

\author[0000-0002-9017-3567]{Anna Y. Q. Ho}
\affiliation{Department of Astronomy, Cornell University, Ithaca, NY 14850}

\author[0000-0002-5936-1156]{Assaf Horesh}
\affiliation{Racah Institute of Physics, The Hebrew University of Jerusalem, Jerusalem 91904, Israel}

\author[0000-0002-4670-7509]{Brian D. Metzger}
\affiliation{Department of Physics and Columbia Astrophysics Laboratory, Columbia University, New York, NY 10027, USA} 
\affil{Center for Computational Astrophysics, Flatiron Institute, 162 5th Ave, New York, NY 10010, USA} 

\author[0000-0001-8472-1996]{Daniel A. Perley}
\affiliation{Astrophysics Research Institute, Liverpool John Moores University, 146 Brownlow Hill, Liverpool L3 5RF, UK}

\author[0000-0002-0592-4152]{Michael Bietenholz}
\affiliation{Department of Physics and Astronomy, York University, Toronto, M3J 1P3, Ontario, Canada}
\affiliation{SARAO/Hartebeesthoek Radio Observatory, PO Box 443, Krugersdorp 1740, South Africa}

\author[0000-0002-3673-0668]{Peter Craig}
\affiliation{Center for Data Intensive and Time Domain Astronomy, Department of Physics and Astronomy, Michigan State University, East Lansing, MI 48824, USA}

\author[0000-0003-0685-3621]{Mark~Gurwell}
\affiliation{Center for Astrophysics \textbar{} Harvard \& Smithsonian, 60 Garden Street, Cambridge, MA 02138-1516, USA}

\author[0000-0002-3490-146X]{Garrett K.~Keating}
\affiliation{Center for Astrophysics \textbar{} Harvard \& Smithsonian, 60 Garden Street, Cambridge, MA 02138-1516, USA}

\author[0000-0002-2249-0595]{Natalie LeBaron}
\affiliation{Department of Astronomy, University of California, Berkeley, CA 94720-3411, USA}
\affiliation{Berkeley Center for Multi-messenger Research on Astrophysical Transients and Outreach (Multi-RAPTOR), University of California, Berkeley, CA 94720-3411, USA}

\author[0000-0001-6857-018X]{Kevin S. McCarthy}
\affiliation{Kavli Institute for the Physics and Mathematics of the Universe (WPI), The University of Tokyo Institutes for Advanced Study, The University of Tokyo, Kashiwa, Chiba, 277-8583, Japan}

\author[0000-0003-0216-8053]{Giulia Migliori} 
\affiliation{INAF -- Istituto di Radioastronomia, Via Gobetti 101, I-40129, Bologna, Italy} 

\author[0000-0001-8426-5732]{Jean Somalwar}
\affiliation{Department of Astronomy, University of California, Berkeley, CA 94720-3411, USA}
\affiliation{Berkeley Center for Multi-messenger Research on Astrophysical Transients and Outreach (Multi-RAPTOR), University of California, Berkeley, CA 94720-3411, USA}
\affiliation{Kavli Institute for Particle Astrophysics and Cosmology, Stanford, CA 94305, USA}

\author[0000-0003-2686-9241]{Daniel Stern}
\affiliation{Jet Propulsion Laboratory, California Institute of Technology, Pasadena, CA 91109, USA}

\author[0000-0002-1407-7944]{Ramprasad~Rao}
\affiliation{Center for Astrophysics \textbar{} Harvard \& Smithsonian, 60 Garden Street, Cambridge, MA 02138-1516, USA}

\author[0000-0002-1468-9668]{Jay Strader}
\affiliation{Center for Data Intensive and Time Domain Astronomy, Department of Physics and Astronomy, Michigan State University, East Lansing, MI 48824, USA}

\author[0000-0001-8023-4912]{Huei Sears}
\affiliation{Department of Physics and Astronomy, Rutgers, the State University of New Jersey, 136 Frelinghuysen Road, Piscataway, NJ 08854-8019, USA}

\author[0000-0003-0466-3779]{Itai Sfaradi}
\affiliation{Department of Astronomy, University of California, Berkeley, CA 94720-3411, USA}
\affiliation{Berkeley Center for Multi-messenger Research on Astrophysical Transients and Outreach (Multi-RAPTOR), University of California, Berkeley, CA 94720-3411, USA}

\author[0000-0003-1814-8620]{Ryan Urquhart}
\affiliation{Center for Data Intensive and Time Domain Astronomy, Department of Physics and Astronomy, Michigan State University, East Lansing, MI 48824, USA}

\author[0009-0002-4843-2913]{Eli Wiston}
\affiliation{Department of Astronomy, University of California, Berkeley, CA 94720-3411, USA}
\affiliation{Berkeley Center for Multi-messenger Research on Astrophysical Transients and Outreach (Multi-RAPTOR), University of California, Berkeley, CA 94720-3411, USA}

\begin{abstract}
We present comprehensive multiwavelength (X-ray, UV, optical, and radio) observations of the Luminous Fast Blue Optical Transient (LFBOT) AT\,2024qfm. At $z = 0.227$ and reaching  $L_{\rm x}\approx 10^{44}\,\rm{erg\,s^{-1}}$,  AT\,2024qfm is the most X-ray luminous LFBOT at $\delta t < 20$\,days, and the least luminous at mm wavelengths. The X-ray emission shows a rapid ($\propto t^{-6}$) decay in the first $\approx 6$\,days followed by a brightening at $\delta t \approx 15$\,days that is accompanied by significant spectral hardening.  
Similarly to other LFBOTs, optical spectra remain featureless, and the UV-optical emission is dominated by a hot thermal continuum with  temperature $T> 10000$\,K until $\delta t \approx 14$\,days. AT\,2024qfm shows luminous radio emission 
that is broadly consistent with the interaction of a fast ($\Gamma \beta c \approx 0.2$\,c) shock wave with a dense medium ($n_{\rm CSM} \approx 6 \times 10^{4}\, \rm cm^{-3}$ at $R \approx 1.3 \times 10^{16}$\,cm). Interestingly, the radio spectra do not evolve between $\delta t \approx 40-58$\,days, indicating multiple outflow components shocking the surrounding medium. Peculiar to AT2024qfm is that  this LFBOT reaches 
$L_{\rm x} \approx L_{\rm opt}$ at much earlier times ($\delta t \approx 7$\,days) than observed for AT\,2018cow and AT\,2024wpp, suggesting less efficient re-processing of X-rays. Finally, a compilation of the X-ray spectral properties of all known LFBOTs shows that they consistently exhibit harder X-ray spectra than GRB afterglows, indicating a different emission mechanism and/or regimes between the two source classes.
\end{abstract}


\section{Introduction} 
\label{sec:intro}
Fast Blue Optical Transients (FBOTs) are characterized by rapid rise times ($t_{\rm rise} \lesssim 10\, \rm days$) to peak brightness, blue and featureless optical spectra, and high peak luminosities  \citep[$L_{\rm pk} \approx 10^{42}-10^{45}\, \rm erg\,\rm s^{-1}$;][]{Drout2014,Pursiainen2018,Ho2023-LFBOTs-rates,Toshikage2024}. The most luminous ($L_{\rm pk} \gtrsim 10^{43}\, \rm erg\,s^{-1}$) FBOTs show bright X-ray and radio emission and  are commonly referred to as LFBOTs \citep{Margutti2019-cow,Ho2019-cow} or ``cow-like" events after the prototypical event AT2018cow. LFBOTs are exceptionally rare with an intrinsic rate of $0.001\%-0.01\%$ of core-collapse Supernovae (CCSNe) rate \citep{Perley2026-wpp}, marking them as the end points of an extremely rare channel of stellar evolution. 

In the past 10 years, detailed X-ray and radio observations were obtained for only 11 LFBOTs: AT\,2018cow \citep{Ho2019-cow,Margutti2019-cow,Nayana2021}, CSS161010 \citep{coppejans2020}, AT\,2018lug \citep{Ho2020-AT2018lug}, AT\,2020xnd \citep{Bright2021-xnd,Ho2022-xnd}, AT\,2020mrf \citep{Yao2021-mrf}, AT\,2022tsd \citep{Ho2023-at2022tsd}, AT\,2023fhn \citep{Chrimes2024a-fhn,Chrimes2024b-fhn,Sevilla2026}, AT\,2022adfc, AT\,2023hkw, AT\,2023vth \citep{Sevilla2026}, and AT\,2024wpp \citep{Pursiainen2025-wpp,LeBaron2025,Nayana2025,Perley2026-wpp}. Significant advance has been made on our understanding of LFBOTs from these studies (see \citealt{Ho2026-LFBOT-review} for a review), that we detail below. 

Radio emission from LFBOTs is understood to be synchrotron radiation produced by the interaction of fast-moving ejecta with a dense circumstellar medium \citep[CSM;][]{Margutti2019-cow, Ho2019-cow}. The outflow velocities inferred from synchrotron spectral modeling span non-relativistic \citep[$\Gamma \beta c \approx0.1c$ in AT\,2018cow;][]{Margutti2019-cow,Ho2019-cow,Nayana2021} to mildly-relativistic \citep[$\Gamma \beta c \gtrsim 0.56c$ in CSS161010;][]{coppejans2020} values. The X-ray emission exhibits significant variability with luminosities comparable to those of Gamma-Ray Bursts \citep[GRBs;][]{Margutti2019-cow,RiveraSandoval2018,    coppejans2020, Ho2022-xnd, Bright2021-xnd, Yao2021-mrf,Ho2023-at2022tsd, Nayana2025}. Broad-band X-ray spectral studies of AT\,2018cow \citep{Margutti2019-cow} and AT\,2024wpp \citep{Nayana2025} revealed a transient Compton hump spectral component interpreted as the result of a central high-energy source viewed through asymmetric, expanding ejecta. 

The UV-optical-emission (hereafter $L_{\rm opt}$) has been hypothesized to be dominated by the partial reprocessing of the inner X-ray source. A striking observational signature that supports this physical picture is the coupled evolution of the optical and X-ray luminosities ($L_{\rm x}$). \cite{LeBaron2025} presented the temporal evolution of $L_{\rm x}/L_{\rm opt}$ for AT\,2024wpp and AT\,2018cow, showing that the ratio increased from $\approx 0.01$ near optical peak to $\approx 1$ at late-times. This evolution has been attributed to a decrease in the effective opacity of the reprocessing layer, allowing an increasing fraction of the inner high-energy emission to escape directly. Proposed models differ in the nature and location of the reprocessing medium: the ejecta surrounding a central engine \cite{Metzger22FBOT} or the cool downstream behind a fast-cooling shock propagating into a dense CSM \citep{Taya2026}. 

Minute-timescale optical flares have been detected from AT\,2022tsd at $\delta t \approx 30-120$\,days post-discovery, indicating the flare emitting region to be at least near-relativistic \citep{Ho2023-at2022tsd}. Persistent, slowly fading UV emission \citep{Sun2022,Chen2023,Inkenhaag2023,Inkenhaag2025} has been detected at the position of AT\,2018cow at $\delta t \approx 2-5.4$\, years post-explosion together with an X-ray source $\delta t \approx 0.5-2.7$\,years \citep{Migliori24}. These late-time UV and X-ray detections have been interpreted as radiation from an accretion disc. Thus, there are multiple observational signatures from the existing sample of LFBOTs that strongly indicate the presence of a central engine, likely in the form of accretion.

Several astrophysical progenitor scenarios have been proposed to explain LFBOTs. \cite{Metzger22FBOT} proposed a model in which a merger-induced tidal disruption of a Wolf-Rayet (WR) star by a compact object powers the transient and associated multi-wavelength emission. A similar physical scenario in a different astrophysical setting is presented by \cite{Tsuna2025}, where a natal kick imparted to the newly formed compact object during a stripped-envelope supernova directs it toward the companion star, leading to its tidal disruption. More recently, a detailed model of merger driven tidal disruption in a binary system followed by a delayed dynamical instability which could produce LFBOTs has been presented in \cite{Klencki2026}. This model naturally explains the broadly similar CSM densities inferred for LFBOTs \citep{Nayana2025,Sevilla2026} as a consequence of mass loss during the prolonged phase of stable mass transfer preceding the merger.

Central-engine models are generally favored for the best studied LFBOTs \citep[AT\,2018cow, AT\,2024wpp;][]{Margutti2019-cow, Nayana2025, Perley2026-wpp}. However, other progenitor models cannot be ruled out. For instance, \cite{Taya2026} proposed a shock-CSM interaction model to reproduce the early multi-wavelength emission from AT\,2018cow. While this model does not require a central-engine to explain the emission on month-long timescales, it does not rule out a central engine contributing to the late-time UV/optical emission. With the limited number of well observed LFBOTs, there is currently no clear consensus on their progenitor scenario. Expanding the sample of LFBOTs with sensitive multi-wavelength observations is therefore essential to understand their intrinsic nature. 

In this paper, we present detailed multiwavelength observations and modeling of the LFBOT AT\,2024qfm (also known as ZTF24aaxhxhf). The paper is organized as follows: we provide background on AT\,2024qfm and prior published results in \S \ref{sec:AT2024qfm-background}. We present UV-optical, X-ray, and radio observations in \S \ref{sec:UV-optical-observations}, \S \ref{Sec:Xraydata}, and \S \ref{Sec:Radio-observations}, respectively. The X-ray spectral analysis and inferences are in \S \ref{Sec:Xraymodeling} and the radio modeling is detailed in \S \ref{Sec:radio-modeling}. We place AT\,2024qfm in the context of other multi-wavelength LFBOTs in \S \ref{sec:comparison}. We summarize our findings in \S \ref{Sec:summary}.

\section{AT\,2024\MakeLowercase{qfm}}
\label{sec:AT2024qfm-background}
AT\,2024qfm was first detected by the Zwicky Transient Facility (ZTF) on 2024-07-24 UT (MJD 60515.41) at RA: 23:21:23.439, Dec = +11:56:32.21 (J2000) with an $r$-band magnitude of 19.57$\pm$0.22 \citep{Perley2024-qfm-optical-followup,Forster2024-qfm-discoveryTNS,Sevilla2026,Fulton2026}. 
Constraining non-detections\footnote{https://alerce.online/object/ZTF24aaxhxhf} were reported on MJD 60513.40 \citep{Sevilla2026,Fulton2026} and we use the midtime between MJD 60513.40 and 60515.41 as the time of first light, i.e., MJD 60514.41. The source was identified to be fast-evolving with a $\approx (2-4)$\,days to rise to peak brightness \citep{Fulton2024-qfm-rapid-evolution}. 

Optical spectroscopic observations with the Gemini Multi-Object Spectrograph (GMOS) on Gemini North on 2024-08-01 ($\delta t \approx 8.6$\,days) showed a blue and featureless spectrum similar to that of AT\,2018cow \citep{Prentice2018-cow-paper,Perley2019,Margutti2019-cow}. The host galaxy spectrum showed H$\alpha$, H$\beta$, [O III], [N II] and [S II] spectral lines, all at a common redshift of $z=0.227$ \citep{Gillanders2024-qfm-redshift,Fulton2024-qfm-rapid-evolution}. The redshift was further confirmed using a Gemini-North GMOS spectrum taken on 2024-08-02 \citep[$\delta t \approx 9.6$\,days;][]{Sevilla2026}. At this redshift the absolute $g$-band magnitude of \at{} at peak was $M_{\rm g} = -21.0$, similar to that of AT\,2018cow \citep{Fulton2026}. The VLA observed the position of \at{} on 2024-08-02 ($\delta t = 9.6$\,days) and did not detect radio emission at 10\,GHz to a 3-sigma flux density limit of $15\, \mu \rm Jy\,\rm beam^{-1}$ \citep{Ho2024-qfm-radio}. X-ray emission was detected from the \textit{Swift}-XRT observations on 2024-08-01 ($\delta t = 8.6$\,days) with a flux of $\approx 5 \times 10^{-13}\, \rm erg\,s^{-1}\,cm^{-2}$ in the $0.3-10$\,keV band \citep{Margutti2024-qfm-Swift}. Follow-up UV-optical photometry and spectroscopy of \at{} is reported by \cite{Sevilla2026} and \cite{Fulton2026}. Taken together, these observations show rapid light curve evolution, persistent blue colors, and featureless spectra, characteristic of LFBOTs, establishing \at{} as a member of this class. 

 We adopt a Lambda cold dark matter cosmology, $\Lambda$CDM $H_0=67.4\,\rm{km\,s^{-1}Mpc^{-1}}$, $\Omega_m=0.315$, $\Omega_{\Lambda}=0.685$ \citep{Cosmology20}. For these parameters the redshift $z=0.227$ corresponds to a luminosity distance of 1170.4 Mpc. We refer to our times with respect to MJD 60514.41, and all times are in the observer frame unless specified.

\section{Optical and UV Observations}
\label{sec:UV-optical-observations}
\subsection{Swift-UVOT (UV) and optical photometry}
\label{subsec:UV}
The Neil Gehrels \emph{Swift} Observatory \citep{Gehrels04} observations of \at{} were acquired in the time range $\delta t=$8.95-134\,days (ObsIDs 16748 and 16763; PIs Chen, Margutti). The UV Optical Telescope (UVOT) onboard \emph{Swift} \citep{Roming05} observed in the UV bands only ($w1$, $w2$, and $m2$). We reduced and analyzed the UVOT data following standard practice as described in \cite{LeBaron2025}, co-adding individual exposures to increase the signal-to-noise ratio. The host galaxy flux is negligible at early epochs $\delta t \lesssim 16$\,days. We use the latest observations at $\delta t \approx 134$\,days, when the transient emission is negligible, as templates for host subtraction. Our final UVOT photometry is reported in Table \ref{Tab:UVOT-photometry}. 

We obtained imaging of AT\,2024qfm with the Goodman High-Throughput Spectrograph on the SOAR 4.1 m telescope on 2024-08-11 in $g$ and $r$ for 2x600s and 3x400s, respectively. We reduced, aligned, and stacked the images using custom routines involving Python and \texttt{lacosmic}\footnote{https://zenodo.org/records/18319010}. We used \texttt{hotpants} \citep{Becker2015-hotpants} to subtract an SDSS image of the field. Using the same SDSS images for magnitude normalization, we performed aperture photometry on the residual images and measured transient magnitudes of $m_{\rm g}$ = 22.24 $\pm$ 0.13 and $m_{\rm r}$ = 22.19 $\pm$ 0.19 AB mag.  

We combine the photometric measurements presented here with those published in \cite{Sevilla2026} to construct the bolometric light curve of \at{}. We account for Galactic extinction along the line of sight by adopting $E(B-V)=0.043$\,mag and $R_{\rm V}=3.1$, using the extinction prescription of \cite{Fitzpatrick1999}, corresponding to $A_{\rm V}=0.134$\,mag. We fit the SEDs where we have \textit{Swift}-UVOT observations. 
The SEDs are well described by a blackbody model, with bolometric luminosities of
$L \approx (4.7,\ 2.4,\ 0.6)\times10^{43}\,\mathrm{erg\,s^{-1}}$
at $\delta t_{\rm rest} = (7.4,\ 9.4,\ 14.2)$ days, respectively. The corresponding temperatures and radii are: $T_{\rm BB} \approx$ 16500, 14200, and 10900\,K and $R_{\rm BB} \approx (9.4, 9.0, 7.7) \times 10^{14}$\,cm, implying average photospheric expansion velocities of $R_{\rm BB}/\delta t_{\rm rest} \approx (0.02-0.05)c$. In addition, we fit blackbody model to the optical photometry of \citep{Fulton2024-qfm-rapid-evolution} at two epochs with sufficient multi-band coverage, $\delta t_{\rm rest} = 6.0$ and 6.7\,days, obtaining $L \approx 3.7$ and $4.3 \times 10^{43}\,\mathrm{erg\,s^{-1}}$, respectively.

\subsection{Optical Spectroscopy}
\label{subsec:optical-spectroscopy}
We obtained optical spectra of AT\,2024qfm during the time interval $\delta t \approx 10-127$\,days at 6 epochs. Three spectra (at $\delta t \approx 13, 17, \, \rm and \, 127$\,days) were obtained using the Keck Low-Resolution Imaging Spectrometer \citep[LRIS;][]{Oke1995} mounted on the 10\,m Keck-I telescope. Three spectra (at $\delta t \approx 10, 11,$ and 19\,days) were obtained using the Double Spectrograph on the Palomar 5\,m telescope \citep[DBSP;][]{Oke1982}. The LRIS observations used a 1$\arcsec$-wide slit with the 400/3400 grism on the blue side, the d560 dichroic beamsplitter, and the 400/8500 grating on the red side to cover the full optical range of 3150--10,300\,\AA. The 127\,d spectrum was taken at the object position with the slit position angle fixed to 68$^{\circ}$ to match that of the 17\,d observation and facilitate accurate removal of the host galaxy contribution under the assumption that the transient had completely faded by that time. Data reduction followed standard procedures, as outlined by \citet{silverman2012}, including telluric correction for the LRIS spectra. 

\begin{figure*} 
 	\centering	\includegraphics[width=0.95\textwidth]{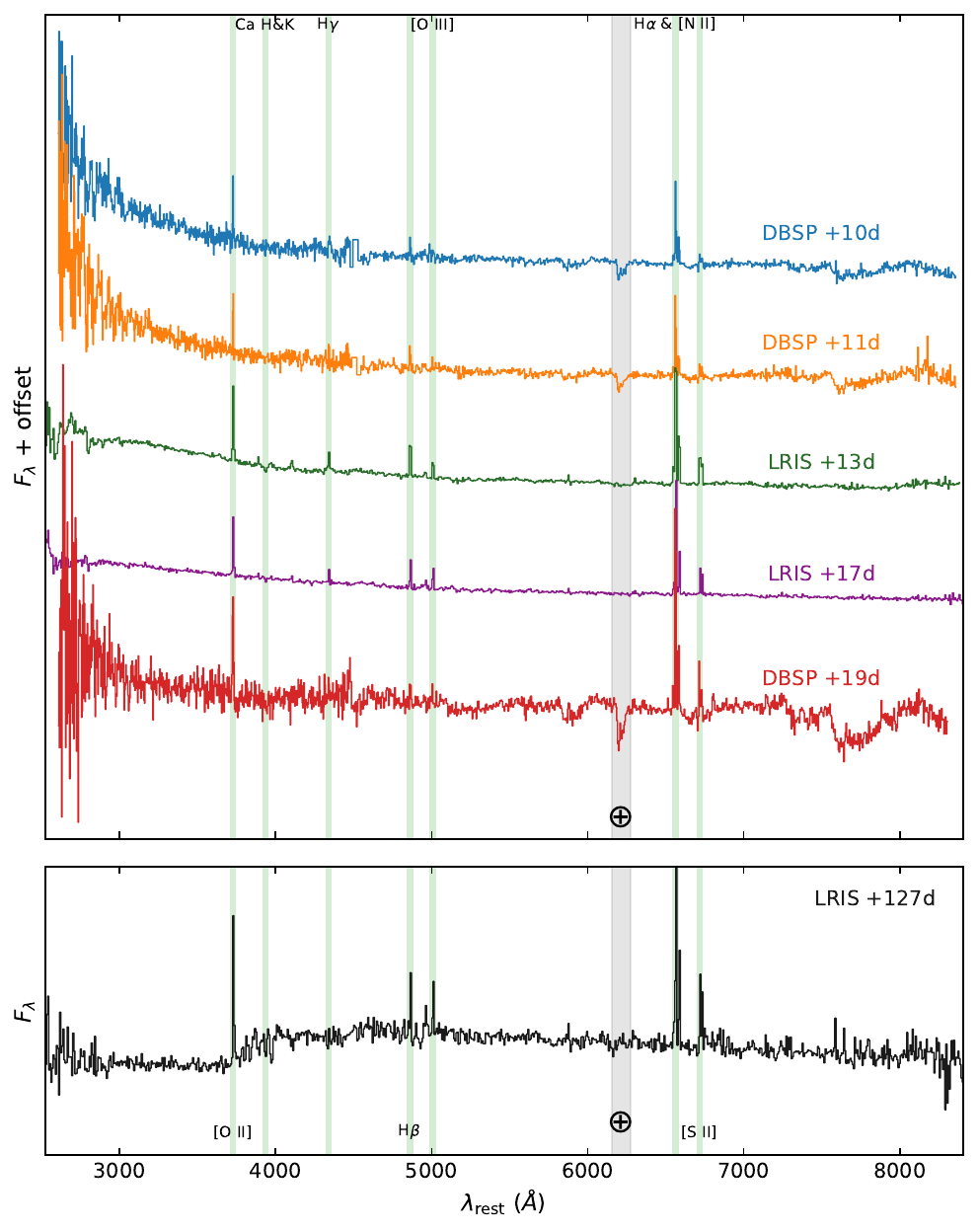} 
  \caption{Upper panel: Optical spectra of \at{} in the time range $\delta t \approx 10-19$\,days obtained with Keck-I/LRIS and Palomar/DBSP. The spectra are plotted with an arbitrary offset. Green shaded regions indicate host galaxy emission lines. The grey shaded region corresponds to the wavelength of telluric absorption bands. Lower panel: The Keck-I/LRIS spectrum obtained at $\delta t \approx 127$\,days. This spectrum is expected to be host dominated as the transient had faded by this epoch. }
 \label{Fig:Optical-spectra}
 \end{figure*}

We used the python package \texttt{prospector} \citep{Johnson2021} that uses the Flexible Stellar Population Synthesis (FSPS) library \citep{Conroy2009} to model the host galaxy's SED. We adopted a parametric star-formation history and a \citet{Calzetti2000} dust attenuation law. Nebular emission was not included in the model, so the resulting template represents only the stellar continuum and absorption features of the host. Following the approach of \citet{Hammerstein2026}, we scaled the best-fit model spectrum to each epoch after matching it to the spectral resolution of each instrument, and subtracted it from the observed spectra. The resulting host-subtracted spectra are presented in Figure \ref{Fig:Optical-spectra}. The spectra are largely featureless and blue without any evidence for transient features, similar to the spectra seen in other LFBOTs \citep{Perley2019,LeBaron2025,Perley2026-wpp}. Since the host emission lines are not part of the model, the narrow emission lines from the host galaxy remain in the subtracted spectra.

\cite{Fulton2026} presented optical spectra of \at{} obtained with Gemini GMOS-N ($\delta t \approx 9$\,days) and ALFOSC on the Nordic optical telescope ($\delta t \approx 9$\,days). Additional spectra from Gemini GMOS-N ($\delta t \approx 10$\,days), Keck LRIS ($\delta t \approx 17$ and 41\,days), and the Binospec instrument on the 6.5m Multiple Mirror
Telescope ($\delta t \approx 39$\,days) have been presented in \cite{Sevilla2026}. Both these studies report optical spectra devoid of any transient spectral features, similar to our conclusion.   

\section{Broad-band X-ray Observations} \label{Sec:Xraydata}
We present soft and hard X-ray observations of \at{} with \emph{Swift}-XRT, \emph{Chandra}, NICER, and \nustar. We note that \at{} was also detected by the Follow-up X-ray Telescope (FXT) on board the Einstein Probe (EP) mission with unabsorbed flux $F_x=6.9^{+6.3}_{-2.3}\times 10^{-14}\, \rm erg\,s^{-1}\,cm^{-2}$  in the 0.5-10 keV band on 2024-08-04, 00:52:31\,UT, which is $\delta t=11.64$\,days \citep{EinsteinProbe}.
\subsection{Swift-XRT (0.3--10 keV)} \label{SubSec:XRTdata}
The X-Ray Telescope (XRT, \citealt{Burrows05}) onboard \emph{Swift} started observing \at{} on  2024-08-01, 08:24:34\,UT until 2024-12-04, 10:25:43\,UT, which is $\delta t=$8.95-134\,days (ObsIDs 16748 and 16763; PIs Chen, Margutti; total exposure time of 24\,ks). The XRT data have been reprocessed with \texttt{HEASoft} v6.36 and corresponding calibration files. We find that the source displays significant temporal variability on time scales of $\sim$days, confirming our preliminary results reported in \cite{Margutti24TNS}.   

We note that a rebinning scheme based on ObsIDs leads to the appearance of a sharp ``spike'' in the count-rate of the source, corresponding to a source detection in Obs ID 00016748007, surrounded by two non-detections in  00016748006 and 00016748008, respectively. However, we note that the detection of Obs ID 00016748007 relies on $N=2$ events that pass the filtering criteria. As these events were acquired over a short exposure time of only $\approx 40$\,s, this means that the interpretation needs to account for the deep Poissonian nature of the signal (and related uncertainties). For example, we note that the observed ``fading'' of the source between  Obs ID 00016748007 and 00016748008 is consistent with the chance fluctuation of a constant source at the $2$\% level, and thus it is not statistically  significant. 
Below we use the 0.3--10 keV count-rate light-curve rebinned to have a minimum number of 5 counts per bin following standard procedures (\citealt{Evans09,Margutti13}).

We extracted three spectra in the time intervals $\delta t=8.95-9.23$\,days, $\delta t=10.07-11.06$\,days (i.e., coordinated with the first \nustar\, observation), and $\delta t=15.14-21.11$\,days. We modeled the spectra using the W-statistics with \texttt{Xspec}: we set the metal abundances to Solar values with \texttt{wilm} adopting the \texttt{tbabs} cross-sections. We find no statistical evidence for intrinsic absorption: we thus adopt $\rm{NH_{int}}=0\,\rm{cm^{-2}}$ and we correct for the Galactic column density along the line of sight, which is  $\rm{NH_{MW}}=4.7\times 10^{20}\,\rm{cm^{-2}}$ \citep{HI4PI}. The detailed flux calibration of the XRT count-rates is based on these spectra and the broad-band X-ray spectral modeling described in \S\ref{Sec:Xraymodeling}.

\subsection{Chandra X-ray Observatory (CXO) (0.3--10 keV)} \label{SubSec:CXOdata}
We obtained six epochs of deep X-ray observations with ACIS-S onboard the CXO (PI Margutti), covering the time range $\delta t=14.62-104.93$\,days. We reduced the \chandra\, ACIS-S data following standard procedures using \texttt{CIAO v4.16} and the associated calibration files. A X-ray source is blindly detected, whose position is consistent with that of \at{} using \texttt{wavdetect} in the first four CXO epochs at $\delta t_{\rm rest} \le 33.3$\,days. For these  epochs, we generated spectra via \texttt{specextract}, defining a source region with a 2\arcsec\, radius and a background region exceeding 35\arcsec\, in radius. The source is no longer detected at $\delta t_{\rm rest} \ge 56.1$\,days. We modeled the spectra with an absorbed power-law model as in \S\ref{SubSec:XRTdata}. The broad-band X-ray spectral modeling is described in \S\ref{Sec:Xraymodeling}. A log of the CXO observations can be found in Table \ref{Tab:Xraylog}.
\subsection{NICER (0.4--15 keV)} \label{SubSec:NICER}
NICER's X-ray Timing Instrument, composed of 52 active Focal-Plane Modules (FPMs), observed AT2024qfm in 53 ObsIDs spanning 2024 July 31 ($\delta t \approx 7.6$\,days) to 2024 October 30 ($\delta t \approx 98.6$\,days). Data were processed using the {\sc nicerdas} software tools, version 2025-09-05\_V015.  Each observation was reduced using {\sc nicerl2} implementing several nonstandard settings in order to reduce particle background and optical loading: South Atlantic Anomaly passages were excluded and data extraction was restricted to intervals with cutoff-rigidity above 3 GeV/c and overshoot rates below 2 s$^{-1}$ FPM$^{-1}$.  Optical loading, which predominantly impacts the low-energy bandpass, was minimized by requiring the undershoot rate $<30$\;s$^{-1}$ FPM$^{-1}$. These values were determined based on an initial exploratory screening of the data.  Spectra were extracted using \texttt{nicerl3-spect}, and the background modeled by fitting with {\sc scorpeon} alongside the source spectral model. Data were analyzed in the energy range 0.4--15\;keV to help ensure the fidelity of the background model, although the primary source sensitivity is $\lesssim$5~keV.  The lower energy bound was shifted higher than the typical 0.25~keV in order to mitigate residual optical loading effects.

\subsection{NuSTAR (3--80 keV)} \label{SubSec:NuSTARdata}
We obtained three epochs of hard X-ray observations with the Nuclear Spectroscopic Telescope Array coordinated with soft X-ray monitoring with \emph{Swift}-XRT or the CXO (PI Margutti).  We processed the data with the \textit{NuSTAR} Data Analysis Software (v2.1.2) and CALDB 20240104, utilizing \texttt{nupipeline} and \texttt{nuproducts} to produce high-level data  products (light-curves and spectra). Source events were selected from a region with a 50\arcsec\, radius. We derived the background estimate from a nearby source-free zone. To ensure data quality, we screened for solar flares and radiation belt passages using both standard diagnostic plots and specialized \texttt{python} scripts. In instances where high background levels were detected, we excluded the affected periods by updating the Good Time Intervals (GTIs) used for data extraction. The source is unambiguously detected via a blind search at the transient coordinates in the first two epochs (i.e., until $\delta t_{\rm rest} = 11.5\,$days). In the third epoch the source is weakly detected at $<20$\,keV in the \nustar\, B-module. A complete log of the \nustar\, observations can be found in Table \ref{Tab:Xraylog}. These  hard X-ray constraints are essential for grounding the broad-band spectral fits presented in \S\ref{Sec:Xraymodeling}.


\section{Broad-band Radio Observations} \label{Sec:Radio-observations}
\subsection{mm-band observations with the SMA and NOEMA } \label{SubSec:SMAdata}
We carried out Submillimeter Array (SMA) observations of \at{} on 2024-08-03 ($\delta t = 11$\,days), which were conducted with a Local Oscillator (LO) frequency of 225.538 GHz, seven antennas in the array, and favorable weather conditions ($\tau_{\rm 225 GHz}\sim0.1$). Data reduction was performed with the COMPASS pipeline (Keating et al., \emph{in prep.}), which applies spectral and temporal flagging of data based on the presence of amplitude outliers and low coherence on calibrator targets. Both J2253+161 and J2327+096 were observed as gain calibrators, with antenna-based amplitude and phase gains were derived, and BL\,Lac was observed as the bandpass calibrator. Flux calibration was performed using the Butler-JPL-Horizons 2012 \citep{2012ALMAM.594....1B} model for Uranus. The data were imaged and deconvolved using the \texttt{CLEAN} algorithm \citep{1974A&AS...15..417H}. Radio emission was not detected with a 3$\sigma$ flux density upper limit of $F_{\nu} < 0.57$ mJy (mean rms = 0.19 mJy/beam) corresponding to a 225.5 GHz luminosity limit of $L_{\nu} < 9.3 \times 10^{29}\, \rm erg\,s^{-1}\,Hz^{-1}$. 

Additional mm-observations of AT\,2024qfm were acquired with the Northern Extended Millimeter Array (NOEMA) at four epochs between 2024-08-05 and September 20 ($\delta t = 12.59 - 58.59$\,days) at 94 GHz (PI: Anna Ho), and have been published in \cite{Sevilla2026}. Radio emission was detected in the first three epochs, but was not detected in the last epoch. Flux density measurements from SMA and NOEMA are presented in Table \ref{Tab:radio-NOEMA}.
\subsection{JVLA} \label{SubSec:VLAdata}
We carried out Jansky Very Large Array (JVLA) observations of AT\,2024qfm from 2024-08-10 ($\delta t = 17.59$\,days) to 2024 December 28 ($\delta t = 117.61$\,days) under project 24B-081 (PI: Nayana) in the  C, X, Ku, and K bands. 3C48 was used as the flux density and bandpass calibrator, and J2330+1100  was used as the phase calibrator. We also include one X-band observation obtained on 2024-08-02 ($\delta t = 9.6$\,days) under project 23B-138 (PI: Anna Ho). We calibrated the data using the VLA calibration pipeline in Common Astronomy Software Applications \citep[CASA;][]{CASAteam2022} version 6.4.1.12 and imaged the field using the task \texttt{TCLEAN} \citep{Offringa2017}. 

Radio emission was not detected in the first two observations ($\delta t = 9.6$ and 17.6\,days) with a 3$\sigma$ flux density limit of $F_{\rm 10\,GHz} < 15\, \mu \rm Jy$ and $F_{15\, \rm GHz} < 15\, \mu \rm Jy$,  respectively. We detected radio emission with a flux density of $F_{\rm 17\, GHz} = 36\pm5\, \mu \rm Jy$ on 2024-08-20 ($\delta t = 27.9$\,days). The CASA task \texttt{IMFIT} was used to measure the flux density of the source. We present radio observation logs and flux density measurements in Table \ref{tab:radio-vla}, and spectral energy distributions (SEDs) in Fig \ref{Fig:radioSED-fit-single-epoch}.

\subsection{AMI-LA}
\label{SubSec:AMI-observations}
The Arcminute Microkelvin Imager-Large Array \citep[AMI-LA,][]{Zwart2008,Hickish2018} is an 8-antenna array, observing a 5 GHz band around a central frequency of 15.5 GHz. Using AMI-LA, we observed the field of AT\,2024qfm as part of an ongoing transient monitoring campaign. The field was first observed on 2024-08-10 ($\delta t = 17.6$\,days), and 18 observations were conducted over the next 2 years.

We reduced, flagged, and calibrated the data using a customized data reduction software for AMI-LA \citep{Perrott2013}. We additionally flagged the data using CASA \citep{CASAteam2022} packages. We produced images of the field using the CASA task \texttt{TCLEAN}. We do not detect radio emission at the position of \at{} in AMI observations. The map RMS at the location of AT\,2024qfm were obtained using the CASA task \texttt{IMSTAT}, and the flux density limits are reported in Table \ref{tab:radio-AMI}.
\section{X-ray Modeling and Inferences} \label{Sec:Xraymodeling}
\subsection{Time-dependent Flux Calibration}\label{SubSec:Xspectra}
We jointly modeled the hard and soft X-ray observations of \at{}\, following the method described in \cite{Nayana2025}. We note that  \at{}\, is clearly detected in the first two \nustar\,  observations, which we use for spectral inference. In the third observation, the source is only faintly recovered in the B module, consistent with the extrapolated soft‑X‑ray signal, and therefore does not contribute additional statistical weight. We modeled the broad-band spectra in  \texttt{Xspec} using the W-statistic, solar abundances with \texttt{wilm} and the \texttt{tbabs} cross-sections.  Differently from ATs 2018cow \citep{Margutti2019-cow} and 2024wpp \citep{Nayana2025}, we do not find statistical evidence for a departure from a simple power-law model, the caveat being the more limited statistical significance of the signal above 10 keV for \at{} compared to the other two events. We also find no evidence for intrinsic absorption and we adopt $\rm{NH_{int}}=0\,\rm{cm^{-2}}$ throughout our analysis. We correct for Galactic neutral hydrogen absorption corresponding to $\rm{NH_{MW}}=4.7\times 10^{20}\,\rm{cm^{-2}}$ \citep{HI4PI}. We report the results of our broad-band spectral analysis in Table \ref{Tab:Xrayspecparameters}, together with five additional epochs of soft X-ray spectra modeling. 

 \startlongtable
\begin{deluxetable*}{ccccccccccc}
\tablecaption{X-ray spectral parameters and inferred fluxes. The single power-law (SPL) model is parametrized as $F_{\nu}\propto \nu^{-\beta}$ where $\beta=\Gamma-1$ and $\Gamma$ is the photon index. }
\tablehead{\colhead{Epoch} &\colhead{Instrument} & \colhead{$\delta t^{a}$} & \colhead{$\Gamma$} & \colhead{$\rm{Log_{10}}(F_x)^{b}$} & \colhead{$\rm{Log_{10}}(F_x)^{b}$}\\
& & (d)     & & ($\rm{cgs}$) & ($\rm{cgs}$)  &
\\
     &  &      &   & [0.3--10 keV] & [0.3--30 keV]& 
 }
\startdata
 1& XRT & 8.95 -- 9.23& $1.98^{+0.25}_{-0.24}$&$-12.34^{+0.08}_{-0.08}$& $-12.32_{-0.08}^{+0.09}$\\
2& XRT & 10.07 --  11.06 & \rdelim\}{2}{*}[\,$1.55^{+0.11}_{-0.12}$ ] & \rdelim\}{2}{*}[\,$-12.50^{+0.03}_{-0.04}$ ]&\rdelim\}{2}{*}[\,$-12.21^{+0.03}_{-0.03}$ ] \\
2 & NuSTAR &9.65 -- 10.72 & \\
3& CXO &14.62 -- 14.89 & \rdelim\}{3}{*}[\,$1.86^{+0.12}_{-0.12}$ ] & \rdelim\}{3}{*}[\,$-12.70^{+0.03}_{-0.03}$ ] & \rdelim\}{3}{*}[\,$-12.68^{+0.03}_{-0.03}$ ]\\
3 &NuSTAR & 13.45 -- 14.85 &\\
3& NICER & 13.23 -- 14.54 &\\
4& XRT$^c$ &15.14 --	21.11 & \rdelim\}{2}{*}[\,$2.50^{+0.13}_{-0.13}$ ] & \rdelim\}{2}{*}[\,$-12.83^{+0.05}_{-0.05}$ ]& \rdelim\}{2}{*}[\,$-12.81^{+0.06}_{-0.06}$ ]\\
4&NICER$^c$ & 14.59 -- 20.86 & \\
5&CXO  & 23.50 -- 23.76 & $1.41^{+0.26}_{-0.26}$ & $-13.07^{+0.06}_{-0.06}$ & $-13.04^{+0.07}_{-0.06}$\\
6&CXO &28.79 --	29.04& $1.38^{+0.37}_{-0.36}$ & $-13.41^{+0.10}_{-0.08}$ & $-13.38^{+0.10}_{-0.08}$\\
7&CXO & 40.72 -- 40.99 & $2.21^{+0.58}_{-0.52}$ &$-13.61^{+0.17}_{-0.09}$
 &  $-13.60^{+0.17}_{-0.09}$\\
\enddata
\tablecomments{$^{\rm{a}}$ Observer frame, with respect to time of first light. \\
$^{\rm{b}}$ Unabsorbed.  \\ $^{\rm{c}}$ For this epoch the statistics are dominated by NICER and we estimate a 0.3 systematic uncertainty for the photon index as a result of the combination of the lack of other high-energy coverage, and a marked dropoff in source-to-background ratio of NICER data above $\approx$3 keV.  
\label{Tab:Xrayspecparameters}}
\end{deluxetable*}

\begin{figure} [b!]
 	\hskip -0.6 cm
    \includegraphics[width=0.5\textwidth]{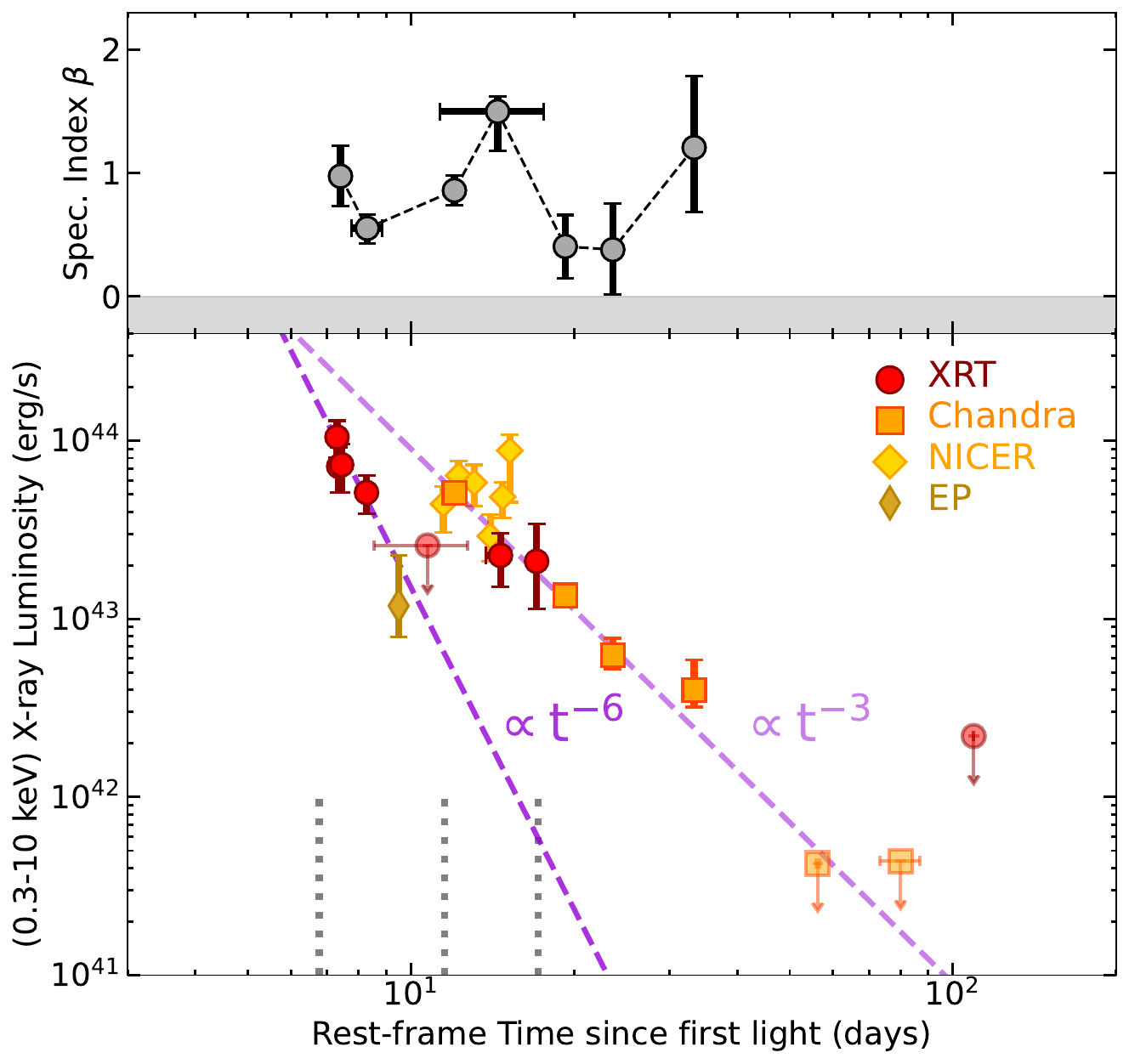} 
  \caption{\emph{Bottom Panel}: 0.3-10 keV X-ray light-curve of \at{}\, as constrained by  \emph{Swift}-XRT (circles), CXO (squares), NICER (diamonds) and  the EP (slim diamond). Dashed purple lines: $L_x\propto t^{-3}$ and  $L_x\propto t^{-6}$ power-law decays to guide the eye. Gray vertical dotted lines: times of acquisition of \nustar\, observations. \emph{Upper Panel:} Spectral index constrained by the broad-band spectral analysis of \S\ref{Sec:Xraymodeling}. The gray shaded area marks the region of a rising $F_{\nu}\propto \nu^{-\beta}$ spectrum. }
 \label{Fig:Xrays24qfm}
 \end{figure}

\subsection{X-ray Spectral  and Temporal Properties}\label{SubSec:Xspectra}
We use the results of our spectral analysis to perform a time-resolved flux calibration of the XRT, NICER and CXO soft X-ray data.  The unabsorbed X-ray light-curve is displayed in Fig. \ref{Fig:Xrays24qfm}:  \at{}\, shows luminous ($L_x\approx 10^{44}\,\rm{erg\,s^{-1}}$) and very rapidly decaying  ($\propto t^{-6}$) X-ray emission in the first $\approx6$\,days, followed by a re-brightening episode at $\approx 15$\,days. Similar to AT\,2024wpp (\citealt{Nayana2025}, their Fig. 1), a pattern emerges where the spectra harden  around the time of the light-curve re-brightening and get softer afterwards. 
The hardest spectra with $F_{\nu}\propto \nu ^{-1/2}$ are consistent with either slow-cooling synchrotron emission, or with fast-cooling free-free emission from a multi-temperature plasma \citep{Wasserman25}, as suggested by \cite{Taya2026} for AT\,2018cow. We further discuss these findings in relation to the optical emission of \at{}\, in \S\ref{SubSec:DiscXrayOpt}.

The lack of evidence for photo-electric absorption is consistent with the large X-ray luminosity of the transient, and the resulting photoionization-induced ``transparency'' of the surrounding medium (e.g., \citealt{Taya26-Xrayabsorption}). Recasting Eq. 8 of \cite{Taya26-Xrayabsorption} following Eq. 5 of \cite{Perley2026-wpp}, the lack of evidence for photo-electric absorption constrains the environment number density as:\footnote{The constraint below applies to ionizing spectra $L_{\nu}\propto \nu^{\beta}$ with $-2 \le \beta \le0$ and a medium density profile with the majority of the mass  concentrated near the inner radius \citep{Taya26-Xrayabsorption}.  }
\begin{equation}
n \ll (3\times 10^{10}\,\mathrm{cm^{-3}})\,\, L_{x,44}^{1/2} r_{15}^{-3/2} \Big (\frac{Z}{Z_{\odot}}\Big )^{-1/2},
\end{equation}
where $L_{x,44}\equiv L_x/10^{44}\,\rm{erg\,s^{-1}}$ (soft X-ray luminosity), $r_{15}\equiv r/10^{15}\,\rm{cm}$ (radius of the absorbing medium), and $Z$ is the metallicity of the medium in units of the solar metallicity, $Z_{\odot}$. The other obvious scenario that would lead to undetectable X-ray absorption is that of X-rays propagating through low-density cavities (e.g., a porous medium).

Finally, we address the X-ray temporal variability. At $\approx10$\,d, \at{} shows an episode of major re-brightening of a factor $\lesssim 10$ with a rise-time scale of a few  days. The brightening has a duration of $\approx 10$\,d and resets the overall light-curve temporal decay in a way that is more reminiscent of energy injection rather than plasma instabilities in radiative shocks (see, however, \citealt{Taya2026}). The rapid rise-time implies a compact emitting region  of $R\lesssim 5\times 10^{15}\,\rm{cm}$, assuming non-relativistic expansion.

\section{Radio Modeling and Inferences}
\label{Sec:radio-modeling}
\subsection{General Considerations}
\label{SubSec:radio-general}
 Radio emission from LFBOTs is understood to be synchrotron radiation arising due to the interaction of fast moving outflow from the explosion with the CSM. The relativistic electrons in the synchrotron emitting source have a power-law distribution of energy $E$: $N(E) \propto N_{0} E^{-p}$ for $E > E_{l}$. Here, $N$ is the number density of relativistic electrons and $E_{l}=\gamma_{\rm m}m_{\rm e}c^{2}$ is the lowest energy of electrons in the power-law distribution, where $\gamma_{\rm m}$ is the minimum Lorentz factor corresponding to a synchrotron characteristic frequency of $\nu_{\rm m}$. The resulting self-absorbed synchrotron spectral energy distribution (SED) is bell-shaped with the optically thick slope of $F_{\nu} \propto \nu^{5/2}$ at $\nu_{\rm m} < \nu < \nu_{a}$, where $\nu_{\rm a}$ is synchrotron self-absorption (SSA) frequency \citep{chevalier1998}. The optically thin slope would be $F_{\nu} \propto \nu^{-(p-1)/2}$ at $\nu_{a} < \nu < \nu_{\rm c}$ and $F_{\nu} \propto \nu^{-p/2}$ at $\nu > \nu_{c}$, where $\nu_{c}$ is the cooling frequency above which electrons efficiently cool over the dynamical timescale. The synchrotron cooling frequency ($\nu_{\rm c}$) is  \citep{RybickiLightman86}:
 \begin{equation}
     \nu_{c} = \frac{18 \pi m_{e} c e}{t^{2} \sigma_{\rm T}^2 B^{3}}
 \end{equation}
 where $\sigma_{\rm T}$ is the Thomson cross section, and $B$ is the post-shock magnetic field, and $t$ is the time since explosion. 
 \begin{figure*} 
 	\centering	\includegraphics[width=0.90\textwidth]{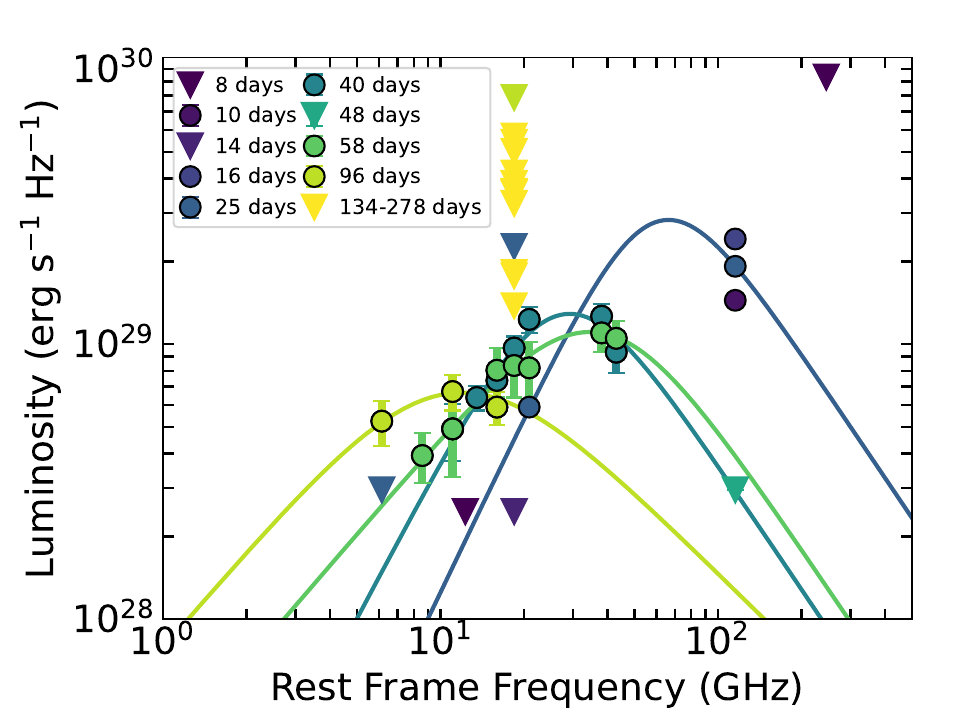} 
  \caption{Radio spectra of \at{} in the time range $\delta t_{\rm rest} \approx 8-278$\,days obtained with our combined JVLA, AMI, NOEMA, and SMA observations. Inverted triangles mark the 3$\sigma$ flux density upper limits. The 15 GHz flux density limits at $\delta t_{\rm rest} \approx 134-278$\,days are from AMI observations. Solid lines denote the best-fit broken power-law models. Note that the SEDs at $\delta t_{\rm rest} \approx 40$ and 58\,days are strikingly similar. }
 \label{Fig:radioSED-fit-single-epoch}
 \end{figure*}
\subsection{Spectral modeling and shock parameters}
\label{subsec:Radio-SED-modeling}
We model SSA spectrum at each epoch as a Broken Power Law (BPL) of the form \citep{DeMarchi2022}
\begin{equation}
    F_{\nu} = F_{\rm pk} \left[ \left( \frac{\nu}{\nu_{\rm pk}} \right)^{\alpha_{1}/s} + \left( \frac{\nu}{\nu_{\rm pk}} \right)^{\alpha_{2}/s} \right]^{s} 
\end{equation}
Here, $\alpha_{1}$ is the optically thin spectral index ($F_{\nu, \rm thin} \propto \nu^{\alpha_{1}}$) and $\alpha_{2}$ is the optically thick spectral index ($F_{\nu,\rm thick} \propto \nu^{\alpha_{2}}$). `$s$' is the smoothing parameter. $\nu_{\rm pk}$ and $F_{\rm pk}$ are the intersection points of the optically thick and thin power laws. The shock radius ($R$) and post-shock magnetic field ($B$) can be constrained from $\nu_{\rm pk}$ and $F_{\rm pk}$ following the \cite{chevalier1998} formulation for $\nu_{\rm m} < \nu_{\rm a} < \nu_{\rm c}$. 
\begin{eqnarray}
\label{eqn:R-C98-slowcooling}
    R=8.8 \times 10^{15} f_{\rm eB}^{-1/19} \left( \frac{f}{0.5}\right)^{-1/19} \left(  \frac{F_{\rm pk}}{\rm Jy}\right)^{9/19} \\ \nonumber
    \times \left( \frac{D}{\rm Mpc} \right)^{18/19} \left( \frac{\nu_{\rm pk}}{5\, \rm GHz} \right)^{-1} \, \rm cm
\end{eqnarray}

\begin{eqnarray}
\label{eqn:B-C98-slowcooling}
    B=0.58 f_{\rm eB}^{-4/19} \left( \frac{f}{0.5} \right)^{-4/19} \left( \frac{F_{\rm pk}}{\rm Jy} \right)^{-2/19} \\ \nonumber \times \left( \frac{D}{\rm Mpc}\right)^{-4/19} \left( \frac{\nu_{\rm pk}}{5\,\rm GHz}\right) \, \rm G
\end{eqnarray}
Here, $D$ denotes the distance to the source and $f_{\rm eB}$ is the ratio of the fraction of post shock energy in relativistic electrons ($\epsilon_{\rm e}$) and amplified magnetic fields ($\epsilon_{\rm B}$); $f_{\rm eB} \equiv \frac{\epsilon_{\rm e}}{\epsilon_{\rm B}}$. The energy density of relativistic electrons is $E_{\rm e} = N_{0} E_{l}^{2-p}/(p-2)$ for $p>2$, and the energy density of magnetic fields is $E_{\rm B} = B^{2}/8\pi$. $f$ denotes the volume filling factor of the synchrotron emitting region. Following \cite{chevalier1998}, we assume $f=0.5$. 

For $\nu_{\rm m} < \nu_{\rm c} < \nu_{a}$ (fast-cooling regime), $R$ and $B$ (see Appendix \ref{Appendix-SSA}) are:

\begin{eqnarray}
\label{eqn:R-fastcooling}
    R=7.76 \times 10^{15} f_{\rm eB}^{-1/13} \left(  \frac{F_{\rm pk}}{\rm Jy}\right)^{6/13} 
     \left( \frac{D}{\rm Mpc} \right)^{12/13} \\ \times \nonumber \left( \frac{f}{0.5} \right)^{-1/13} \left( \frac{\nu_{\rm pk}}{5\, \rm GHz} \right)^{-11/13} \left( \frac{t}{100\, \rm days}\right)^{1/13} \, \rm cm
\end{eqnarray}

\begin{eqnarray}
\label{eqn:B-fastcooling}
    B=0.35 f_{\rm eB}^{-4/13} \left( \frac{F_{\rm pk}}{\rm Jy} \right)^{-2/13} \left( \frac{D}{\rm Mpc}\right)^{-4/13} \\ \nonumber \times \left( \frac{f}{0.5} \right)^{-4/13} \left( \frac{\nu_{\rm pk}}{5\,\rm GHz}\right)^{21/13} \left( \frac{t}{100\, \rm days}\right)^{4/13} \, \rm G
\end{eqnarray}

The CSM density can be estimated from the magnetic field scaling relation; $\frac{B^{2}}{8\pi}=\epsilon_{\rm B}\rho_{\rm CSM}v^{2}$, where $v$ is the mean shock velocity defined as $v = dR/dt$. 

We fit the radio SED at each epoch using a BPL function fixing $s=-1$. Observations at $\delta t_{\rm rest} = 22.7$\,days (17\,GHz), $\delta t_{\rm rest} = 25.1$\,days (5\,GHz), and $\delta t_{\rm rest} = 27.4$\,days (100 GHz) are fit together for an average epoch of $\delta t_{\rm rest} \approx 25.1$\,days. We initialize the optically thin spectral index $\alpha_{1}$ as $\alpha_{1} = -1$ ($p=3$) motivated by the spectral indices seen in other radio bright LFBOTs \citep{Margutti2019-cow,Ho2019-cow,coppejans2020,Nayana2025}. We keep $\nu_{\rm pk}$, $F_{\rm pk}$, and $\alpha_{2}$ as free parameters at $\delta t_{\rm rest} \approx 40$ and 58\,days. At $\delta t_{\rm rest} \approx 25$ and 96\,days only $\nu_{\rm pk}$ and $F_{\rm pk}$ are set as free parameters, fixing $\alpha_{2}$ to be the best-fit values from the closest SED. We use the values of $R$, $B$, and $v$ from these fits, and calculate $\nu_{\rm m}$ and $\nu_{\rm c}$ at each epoch. If $\nu_{\rm c} < \nu_{\rm pk}$, the optically thin slope would be $\alpha_{1}=-1.5$ and we re-fit those SEDs with a BPL using $\alpha_{1}=-1.5$. We calculate $R$ and $B$ using equations \ref{eqn:R-C98-slowcooling} and \ref{eqn:B-C98-slowcooling} in the slow-cooling regime and equations \ref{eqn:R-fastcooling} and \ref{eqn:B-fastcooling} in the fast-cooling regime, respectively. The best-fit parameters and the derived self-consistent values of $R$, $B$, $v$, $n$, and $U$ are given in Table \ref{Tab:radio-shock-parameters}.

\startlongtable
\begin{deluxetable*}{lcccccccccc}
\tablecaption{Best-fit values and shock parameters of \at{} estimated from single-epoch radio SED modeling.}
\tablehead{
\colhead{Time$^{\rm{a}}$} & 
\colhead{$\alpha_{1}$} &
\colhead{$\alpha_{2}$} & \colhead{$\nu_{pk}^{\rm{b}}$} & \colhead{$F_{pk}^{\rm{b}}$}  & \colhead{$R$} & \colhead{$(\Gamma \beta)c$} & \colhead{$B$} & \colhead{$n$} & \colhead{$U$}\\
(d)  &  &  & (GHz) & (mJy) & ($\times 10^{16}$cm) & ($c$) & (Gauss) & (cm$^{-3}$) & ($\times 10^{48}$erg) \\
 }
\startdata
25.1 & $-1.5$ & 1.91 & 65.0$^{+5.5}_{-5.3}$ & 0.316$^{+0.041}_{-0.040}$ 
& $1.32^{+0.09}_{-0.09}$ & $0.203^{+0.014}_{-0.014}$ & $5.66^{+0.76}_{-0.72}$ & $6.25^{+2.85}_{-1.97}\times10^{4}$ & $18.25^{+3.14}_{-3.34}$ \\
40.5 & $-1.5$ & $1.91^{+0.46}_{-0.45}$ & $27.2^{+4.8}_{-2.7}$ & $0.156^{+0.011}_{-0.011}$ 
& $2.06^{+0.26}_{-0.35}$ & $0.196^{+0.025}_{-0.033}$ & $1.79^{+0.33}_{-0.18}$ & $6.70^{+6.81}_{-2.43}\times10^{3}$ & 7.05$^{+1.14}_{-1.39}$ \\
58.4 & $-1.5$ & $1.17^{+0.31}_{-0.22}$ & $37.7^{+7.9}_{-7.4}$  &  $0.134^{+0.014}_{-0.014}$
& $1.49^{+0.34}_{-0.24}$ & $0.098^{+0.022}_{-0.016}$ & $3.48^{+1.28}_{-1.05}$ & $1.00^{+1.6}_{-0.7}\times10^{5}$ & 10.02$^{+1.61}_{-1.38}$ \\
95.9 & $-1.0$ & 1.17 & $11.1^{+3.8}_{-2.4}$  & $0.081^{+0.008}_{-0.007}$ 
& $3.63^{+1.02}_{-0.88}$ & $0.146^{+0.041}_{-0.036}$ & $0.79^{+0.26}_{-0.17}$ & $2.34^{+4.96}_{-1.45}\times10^{3}$ & 7.49$^{+2.37}_{-1.77}$ \\
\enddata
\tablecomments{$^{\rm{a}}$ With respect to date of first light, in the rest frame. \\ $^{\rm{b}}$ $\nu_{\rm pk}$ and $F_{\rm pk}$ are the asymptotic intersection of the optically thick and thin power laws of self-absorbed synchrotron spectrum as prescribed by \cite{chevalier1998}. The parameters are estimated assuming equipartition ($\epsilon_{\rm e}=\epsilon_{\rm B}=0.3$). \\ $^{\rm{c}}$ Mean shock velocity ($\Gamma \beta)c=Rc/t$ 
\label{Tab:radio-shock-parameters}}
\end{deluxetable*}

\begin{figure} 
 	\centering	\includegraphics[width=0.49\textwidth]{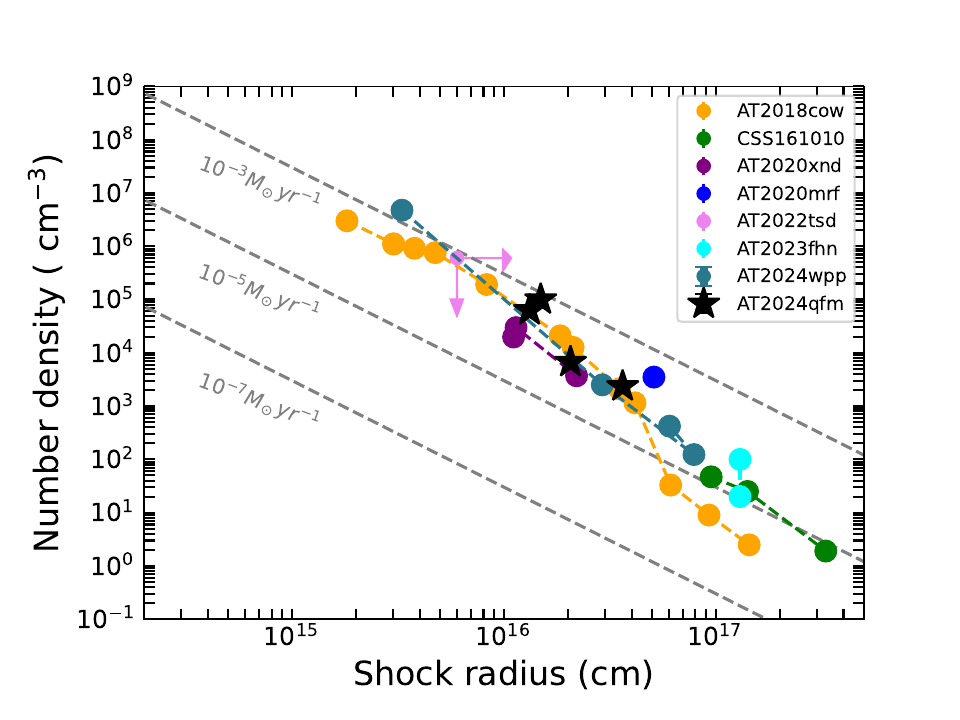} 
  \caption{CSM density profile of AT\,2024qfm (black stars) from radio SED modeling of \S \ref{Sec:radio-modeling} assuming equipartition ($\epsilon_{\rm e}=\epsilon_{\rm B}=0.3$), in the context of the CSM density profiles of other LFBOTs \citep[][and references therein]{Nayana2025}. Dashed grey lines indicate density profiles for constant mass-loss rates for a wind velocity of 1000 km\,s$^{-1}$. LFBOTs show strikingly similar CSM density profiles at comparable radii suggesting common evolutionary pathways for the progenitors.}
 \label{Fig:density-evolution}
 \end{figure}

\subsection{Spectral evolution and  shock parameters}
\label{subsec:R-evolution}
In a standard shock-driven synchrotron emission scenario, the SSA spectra are expected to peak at higher frequencies first and later cascade to lower frequencies if the shock expands and interacts with less dense CSM. While, the overall radio SED evolution of \at{} qualitatively follows this pattern, the spectra between $\delta t_{\rm rest} \approx 40.5$ and 58.4\,days are strikingly similar. The inferred shock radius evolves from $R = 1.32^{+0.09}_{-0.09} \times 10^{16}$\,cm to $R = 2.06^{+0.26}_{-0.35} \times 10^{16}$\,cm between $\delta t \approx 25.1$ and $40.5$\,days and then stays approximately steady until $\delta t \approx 58.4$\,days. The radius again increases to $R = 3.63^{+1.02}_{-0.88} \times 10^{16}$\,cm by $\delta t \approx 95.9$\,days. This SED and radius evolution cannot be explained in a scenario in which a single shock is expanding into a homogeneous CSM. 

We note that for $\epsilon_{\rm e} = \epsilon_{\rm B} = 0.3$, the SED exhibits $\nu_{\rm c} \gtrsim \nu_{\rm pk}$ at $\delta t_{\rm rest} = 40.5$\,days and $\nu_{\rm c} < \nu_{\rm pk}$ at $\delta t_{\rm rest} = 58.4$\,days. If we adopt deviations from equipartition, both SEDs will fall into similar spectral regime of $\nu_{\rm c} \gtrsim \nu_{\rm pk}$ for $\epsilon_{\rm e}/\epsilon_{\rm B} \gtrsim 100$. This would indicate a nearly stalled shock with $R(\delta t_{\rm rest} \approx 40\, \rm days)$ $\approx$ $R(\delta t_{\rm rest} \approx 58\, \rm days$), still inconsistent with the standard scenario.

We consider two alternative scenarios that could explain the observed shock evolution:
(1) Shock interaction with a clumpy CSM, in which the covering factor of the emitting region varies with radius, manifests as an apparent stalling or decrease of the inferred shock radius. 
(2) Emission from multiple outflow components, where a second outflow overtakes the first shock. 
\begin{figure} 
 	\centering	\includegraphics[width=0.49\textwidth]{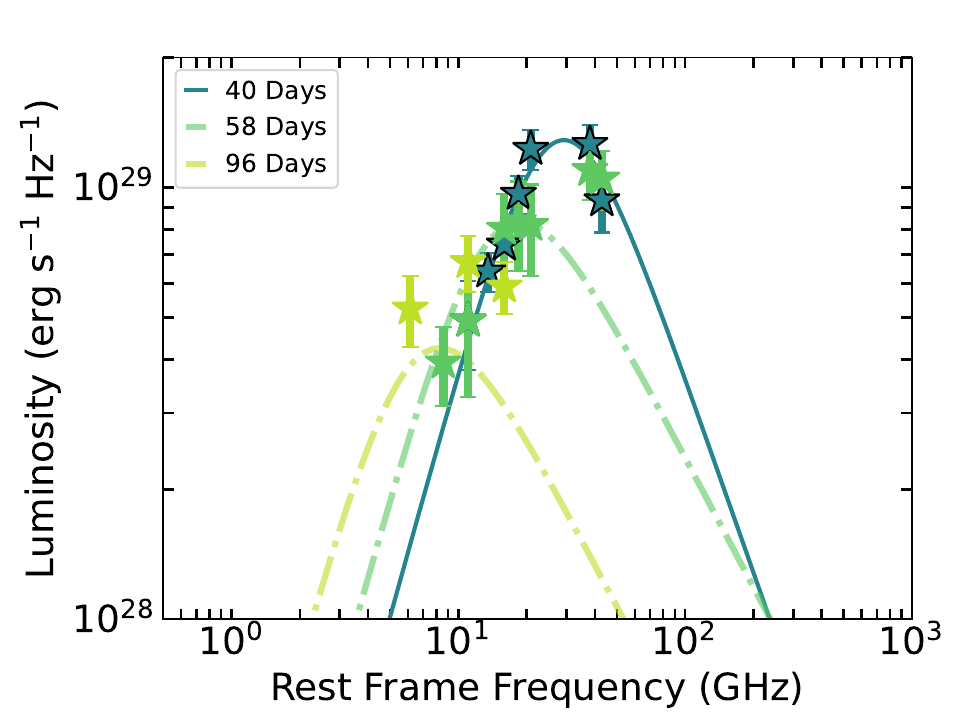} 
  \caption{Evolution of synchrotron SEDs of \at{} at multiple epochs in a single shock scenario. Solid dark green curve represents the best fit BPL model for the spectrum at $\delta t = 40.5$\,days. Green and yellow dotted dashed lines indicate extrapolations of the synchrotron SED to $\delta t = 58$ and 96\,days assuming a constant shock velocity of $\Gamma \beta c \approx 0.196c$ and CSM density profile of $\rho_{\rm CSM} \propto r^{-3}$.}
 \label{Fig:SED-evolution-extrapolated}
 \end{figure}

\begin{figure} 
 	\centering	\includegraphics[width=0.47\textwidth]{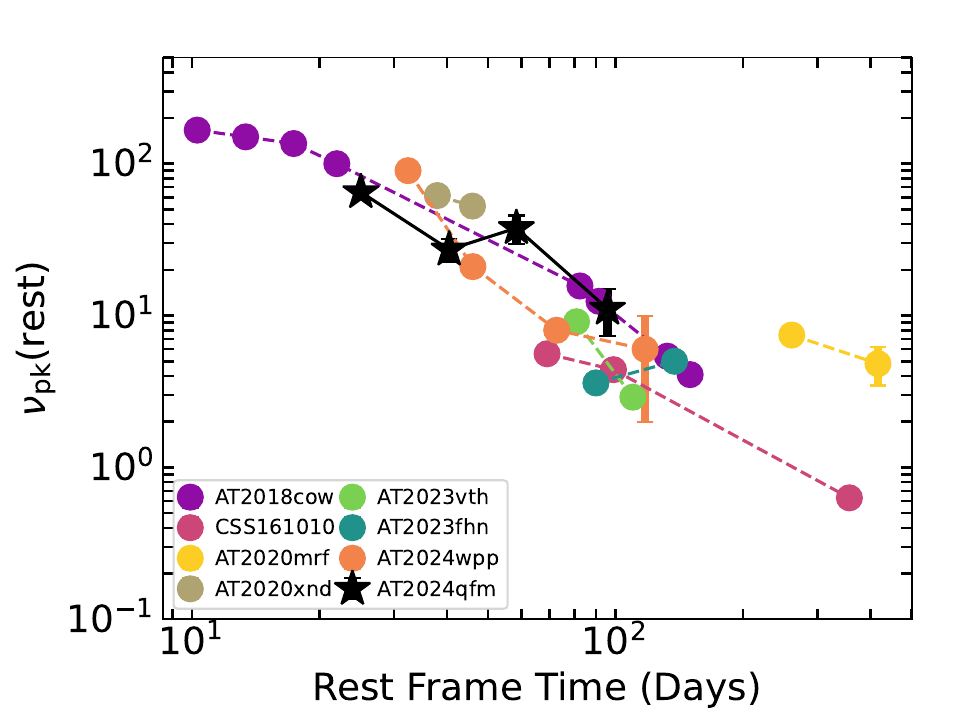} 
  \caption{Evolution of best-fit $\nu_{\rm pk}$ from synchrotron SED modeling of \at{} along with other well-observed LFBOTs \citep{Margutti2019-cow, Ho2019-cow, coppejans2020,Sevilla2026,Chrimes2024b-fhn,Yao2021-mrf,Nayana2025}. Broadly similar evolution of $\nu_{\rm pk}$ across events indicates the universal density profiles seen around LFBOTs.}
 \label{Fig:nup-all-LFBOTs}
 \end{figure}

 \begin{figure} 
 	\centering	\includegraphics[width=0.45\textwidth]{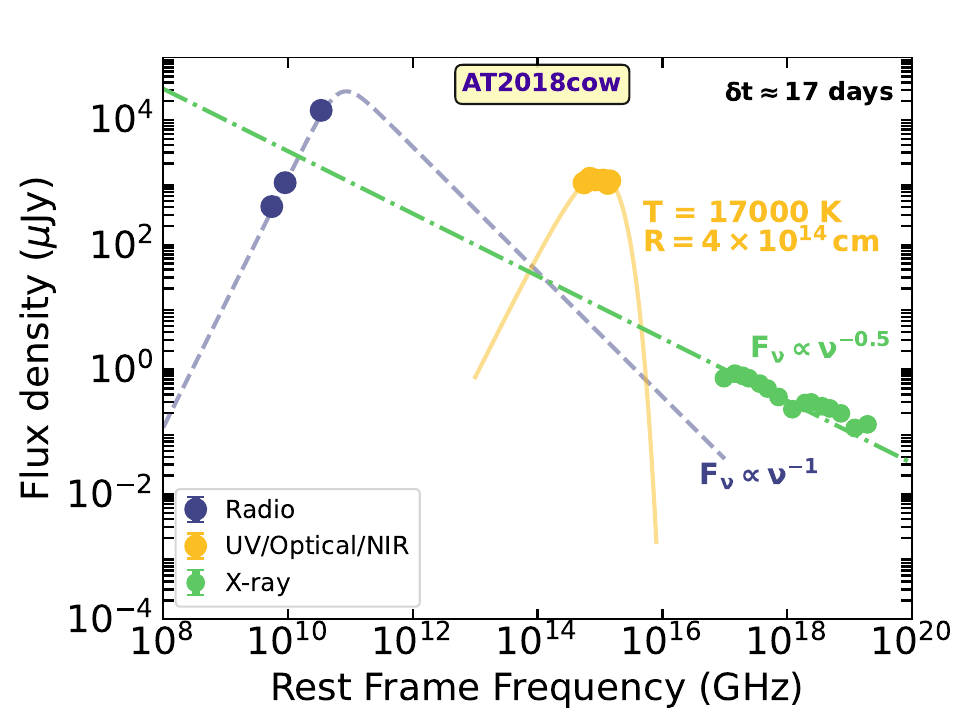} 
    \includegraphics[width=0.45\textwidth]{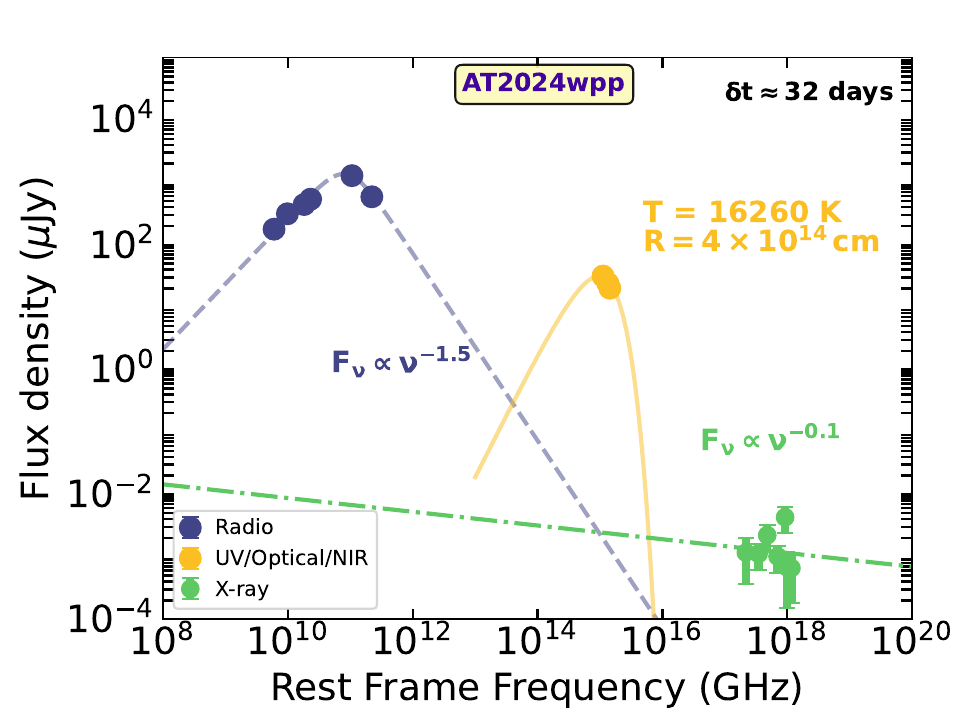} 
    \includegraphics[width=0.45\textwidth]{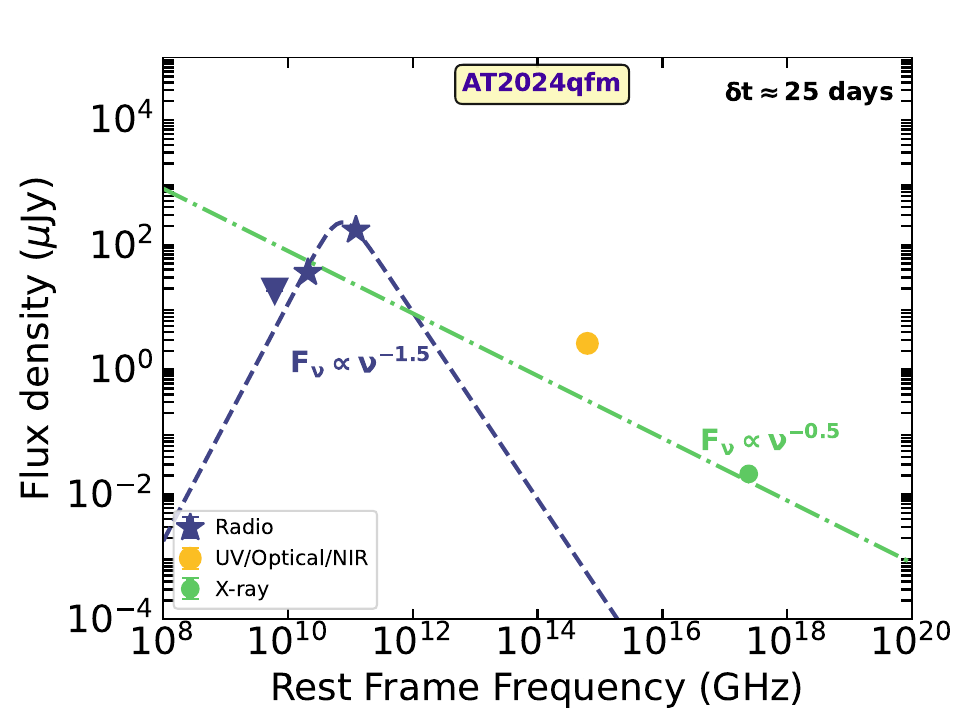} 
  \caption{Broadband spectral energy distributions of LFBOTs AT\,2018cow \citep{Ho2019-cow,Margutti2019-cow}, AT\,2024wpp \citep{Nayana2025}, and AT\,2024qfm (this work). The distinct spectral and temporal evolution observed in the X-ray, optical, and radio bands indicates that the emission in each band originates from a separate component.  }
 \label{Fig:broadband-SEDs}
 \end{figure}

\subsubsection{Varying covering factor}
Depending on the geometry of the emitting region, the value of volume filling factor ($f$) can deviate from the standard assumed value of $f=0.5$ \citep{chevalier1998}. The inferred size of the synchrotron emitting region scales as $R \propto f^{-1/19}$ for $\nu_{m} < \nu_{a} < \nu_{c}$, and $R \propto f^{-1/13}$ for $\nu_{m} < \nu_{c} < \nu_{a}$. Assuming the shock expands at a constant rate of $\Gamma \beta c \approx 0.2c$ at $\delta t \gtrsim 40$\,days, the expected shock radius at $\delta t \approx 58$ is $R \approx 3 \times 10^{16}$\,cm. A covering factor of $f \approx 10^{-6}$ at $\delta t \approx 58.4$\,days  would reproduce a consistent shock evolution. However, the magnetic field strength scales as $B \propto f^{-4/13}$, and would imply $B \approx 200$\,G at $\delta t \approx 58.4$\,days for this covering factor. This $B$ is unrealistically high and would result in extremely large shock energies $\mathcal{O}(10^{53}\,\mathrm{erg}$) suggesting that the assumed combination of $R$, $B$, and covering factor is physically implausible.
\subsubsection{A separate outflow component}
Another plausible physical scenario is the presence of a separate outflow component dominating the synchrotron flux at $\delta t \gtrsim 58.4$\,days. Assuming the first outflow component dominating the synchrotron spectrum on $\delta t =40.5$\,days, we extrapolate this SED to $\delta t =58.4$ and 96\,days for a constant shock velocity of $\Gamma \beta c =0.2c$. We also assume a CSM density profile of $n_{\rm CSM}=n_{0}\left(\frac{r}{r_{0}} \right)^{-3}$, where $n_{0} \approx 4 \times 10^{3}$\,cm$^{-3}$ and $r_{0} \approx 2.3 \times 10^{16}$\, cm from the SED  modeling at $\delta t = 40.5$\,days. The assumed density profile is motivated by the universal behavior of CSM density profiles of all radio-bright LFBOTs with multi-epoch spectral analysis \cite[see Fig. 9 of][]{Nayana2025}. The extrapolated SEDs along with the observed spectral luminosity measurements are shown in Figure \ref{Fig:SED-evolution-extrapolated}. The SEDs at both epochs ($\delta t=58$ and 96\,days) underpredict the observed synchrotron flux densities. If we subtract the extrapolated SEDs, the remaining synchrotron flux density is $F_{\rm pk} = 0.083\pm0.008$\,mJy ($\nu_{\rm pk} = 37.7\pm7.9$\,GHz) at $\delta t =58$\,days and $F_{\rm pk} = 0.047\pm0.004$\,mJy ($\nu_{\rm pk} = 11.1\pm3.8$\,GHz) at $\delta t =96$\,days. 

The physical parameters associated to this second component are the following: $R =(1.1\pm0.2) \times 10^{16}$\,cm, $\Gamma \beta c = (0.07\pm0.02)c$ at $\delta t_{\rm rest} \approx 58$\,days and $R =(2.9\pm0.9) \times 10^{16}$\,cm, $\Gamma \beta c = (0.11\pm0.03)c$ at $\delta t_{\rm rest} \approx 96$\,days. This qualitative analysis indicates an outflow of roughly constant velocity of $\approx 0.1$c moving to an environment that is an order of magnitude denser than the one encountered by the first faster outflow. Thus, the observed radio evolution is consistent with a scenario in which multiple outflow components interact with a complex CSM: a faster component propagating into a lower density CSM that dominates the early emission; and a slower component emerging at late-times from a higher density medium.

\section{AT\,2024\MakeLowercase{qfm} in the phase space of other multi-wavelength LFBOTs}
\label{sec:comparison}
\subsection{Broadband spectral properties}
\label{sec:Xraytoradio-SEDs-FBOTs}
We compile broadband X-ray to radio SEDs of well-observed LFBOTs in Figure \ref{Fig:broadband-SEDs}: AT\,2018cow \citep{Ho2019-cow,Margutti2019-cow}, AT\,2024wpp \citep{LeBaron2025,Nayana2025}, and AT\,2024qfm. While there exist multiple epochs of broad-band observations for AT\,2018cow \citep[see Fig 11 from][]{Margutti2019-cow} and AT\,2024wpp, but the multi-band coverage of AT\,2024qfm is limited. We choose to plot the epoch close to the peak of radio emission for all three events in order to compare similar behavioral regimes. The radio, optical, and X-ray emission from LFBOTs originate from separate emission components, and can have different spectral and temporal evolution \citep[e.g.,][]{Margutti2019-cow,Nayana2025,Ho2023-at2022tsd}. AT\,2024qfm adds to this picture of broadband spectral properties of LFBOTs. The X-ray emission is in clear excess to the power-law extrapolation of the radio SEDs to higher frequencies, which further confirms that in LFBOTs the X-rays are not part of the radio synchrotron spectrum, and are likely to originate from a different emitting region and/or shock. 
While there are differences in relative flux densities and peak frequencies across different bands, a broadly consistent picture of LFBOT broadband behavior is emerging, in which different observing bands trace physically distinct emission components rather than a single underlying spectral shape.
\subsection{Linking X-rays with Optical emission in LFBOTs}
\label{SubSec:DiscXrayOpt}

\begin{figure} 
 	\hskip -0.6 cm
    \includegraphics[width=0.5\textwidth]{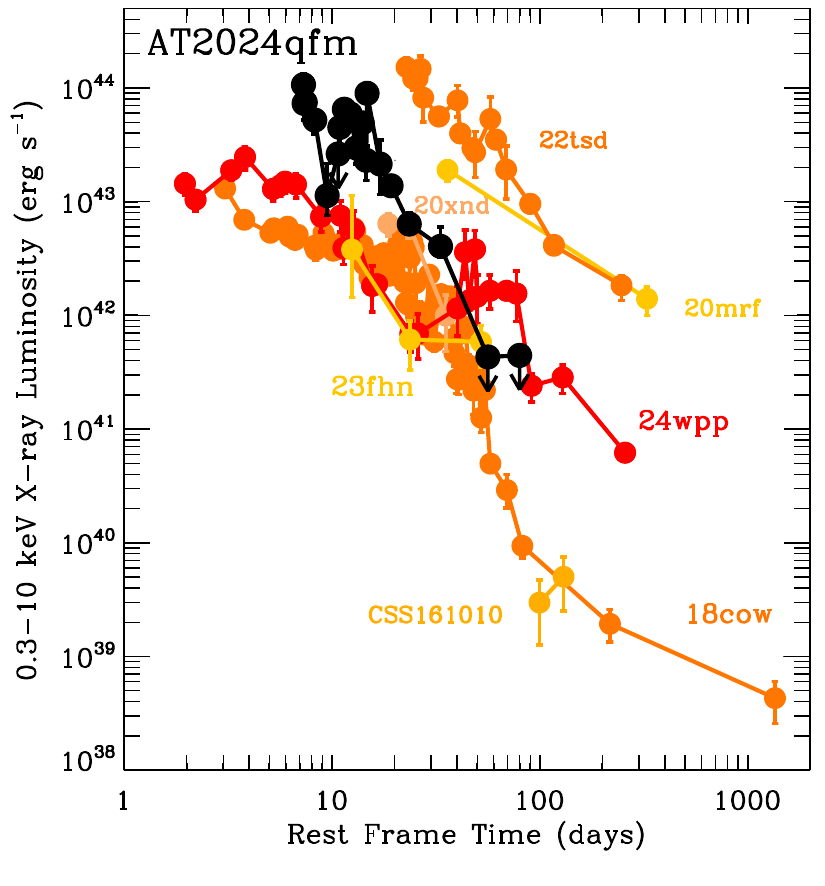} 
  \caption{Soft X-ray emission of \at{}\, (black symbols) in the context of LFBOTs with deep X-ray observations. At $\delta t<20$\,d, \at{}\, is currently the most X-ray luminous LFBOT detected in the soft X-rays. References: CSS161010 \citep{coppejans2020}, ATs 2018cow  \citep{Margutti2019-cow,Migliori24}, 2020xnd \cite{Bright2021-xnd}, 2020mrf \citep{Yao2021-mrf}, 2022tsd \citep{Matthews22tsd}, 2023fhn \citep{Nayana2025}, 2024wpp \citep{Nayana2025}.}
 \label{Fig:XrayFBOTs}
 \end{figure}

\begin{figure*} 
 	\hskip -0.6 cm
    \centering
\includegraphics[width=0.46\textwidth]{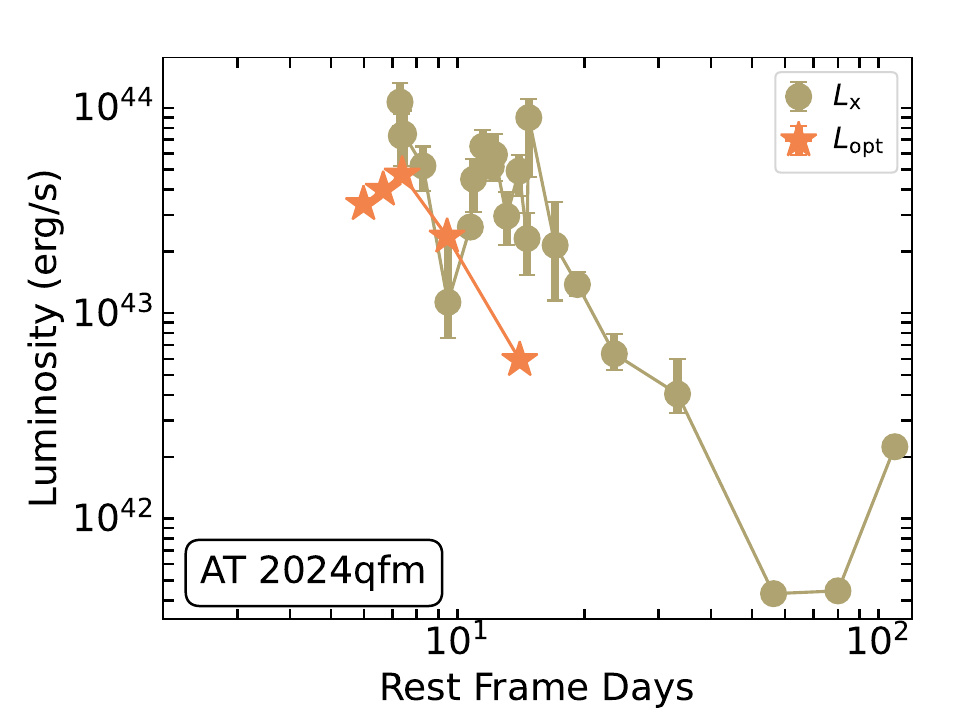}  \includegraphics[width=0.46\textwidth]{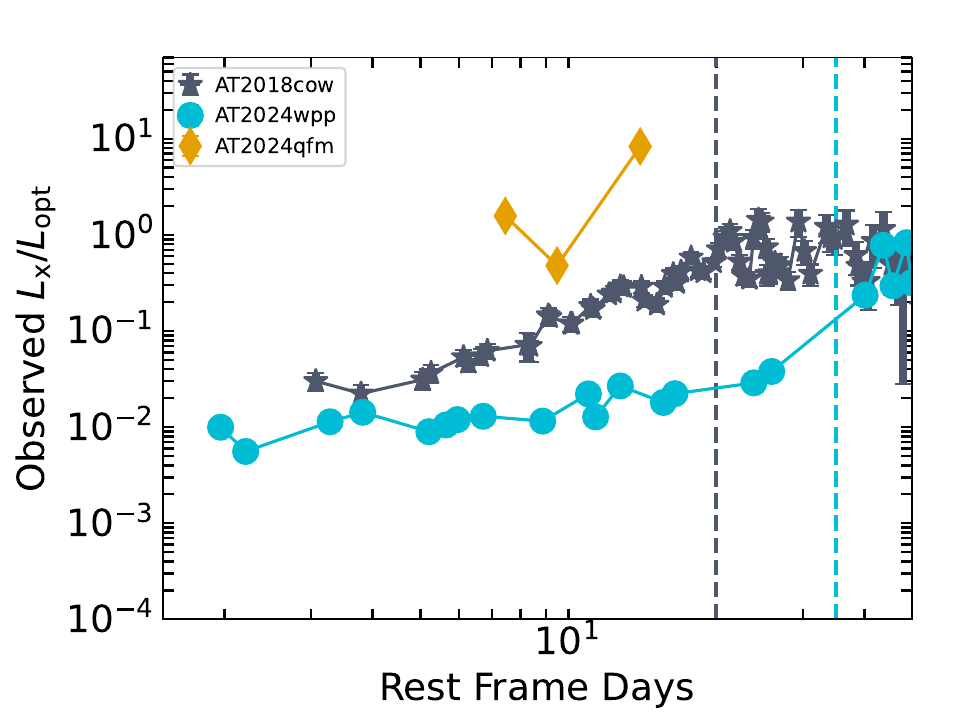} 
  \caption{\emph{Left Panel:} Evolution of 0.3--10 keV X-ray luminosity ($L_{\rm x}$) and UVOIR luminosity ($L_{\rm opt}$) of AT\,2024qfm. The $L_{\rm opt}$ is derived from the blackbody fits. \emph{Right:} Ratio of $L_{\rm x}$ to $L_{\rm opt}$ for \at{} (orange diamonds), AT\,2018cow (grey stars), and AT\,2024wpp (blue circles). Grey and cyan dashed lines indicate the time of appearance of spectral features of a few $\times$ 1000 km\,s$^{-1}$ in AT\,2018cow ($\delta t_{\rm rest} \approx 20$\,days) and AT\,2024wpp ($\delta t_{\rm rest} \approx 35$\,days), respectively. This time also roughly coincides with the time when $L_{\rm x} \approx L_{\rm opt}$ in both events \citep{LeBaron2025}. In \at{}, $L_{\rm x} \approx L_{\rm opt}$ happens at much earlier times ($\delta t_{\rm rest} \approx 7.4$\,days). }
 \label{Fig:Lxray-Lbol}
 \end{figure*}

\begin{figure*} 
 	\hskip -0.6 cm
    \centering
    \includegraphics[width=0.8\textwidth]{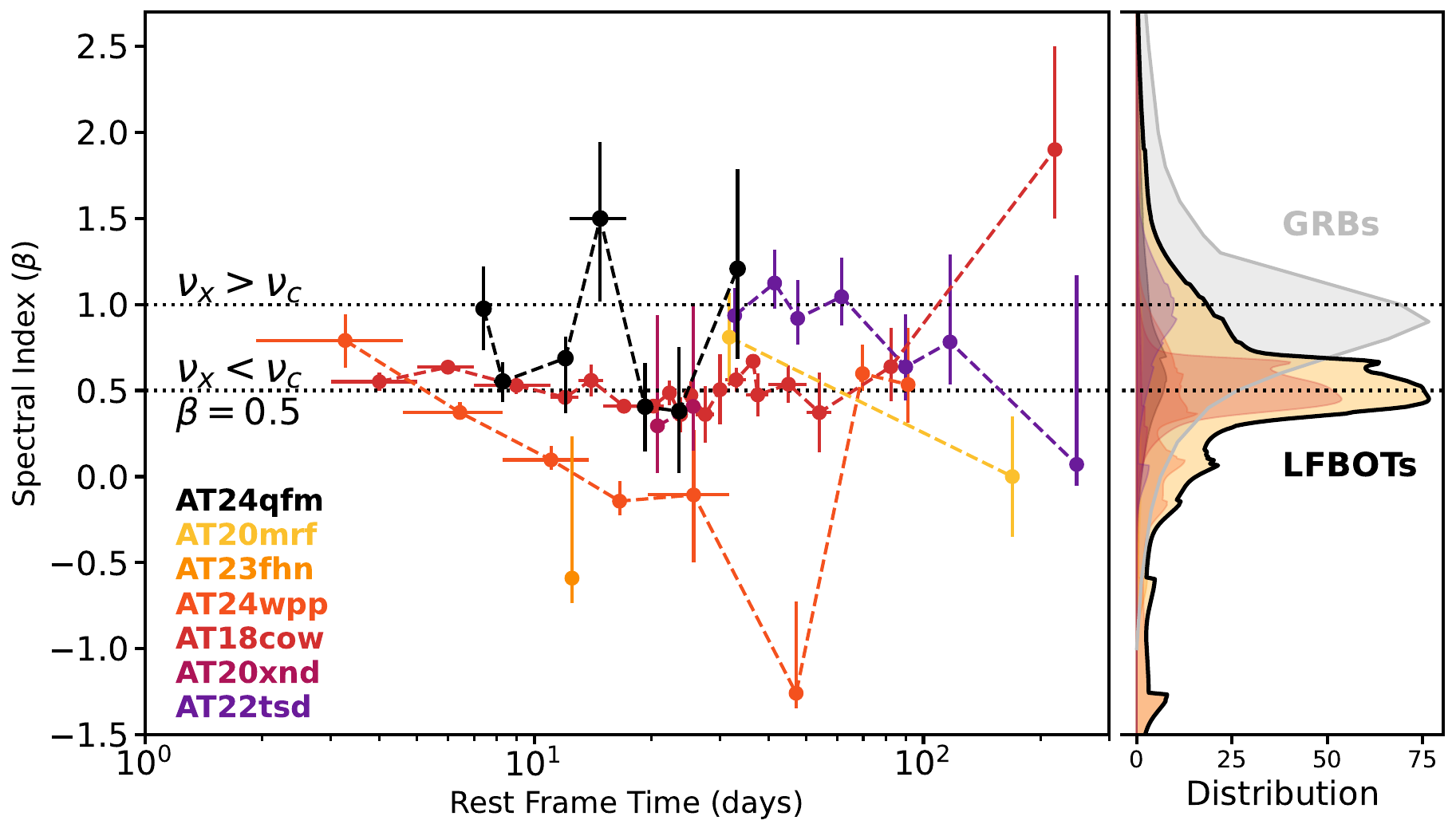} 
  \caption{\emph{Left Panel:} X-ray spectral index evolution $\beta$ (where $F_{\nu}\propto \nu^{-\beta}$ of all LFBOTs for which this parameter can be observationally constrained. Horizontal dotted lines: expected $\beta\approx 1$ value for synchrotron-powered X-rays above the cooling frequency when $F_{\nu}\propto \nu^{-p/2}$ and $p\approx 2$ (upper line), and X-rays below the cooling frequency ($F_{\nu}\propto \nu^{-(p-1)/2}$ and $p\approx 2$, lower line). The soft X-ray spectrum of a free-free cooling multi-temperature plasma also shows  $\beta\approx0.5$ (\citealt{Taya2026}, their Appendix B; \citealt{Wasserman25}), consistent with LFBOTs.    Instead, a single-temperature thermal bremsstrahlung spectrum with temperature above $10$\,keV would manifest as a $\beta \approx 0$ spectrum in this plot. \emph{Right Panel:} resulting distribution for the LFBOT population (yellow shade) vs. GRB afterglows (grey shade). As a population, LFBOTs show  harder X-ray emission than GRBs afterglows, which are known to be powered by the relativistic shock interaction with the medium. References: ATs 2018cow \citep{Margutti2019-cow,Migliori24}; 2020mrf \citep{Yao2021-mrf}; 2020xnd \citep{Bright2021-xnd}; 2022tsd (this work); 2023fhn (this work); 2024wpp \citep{Nayana2025}; 2024qfm (this work). GRBs: distribution of the average photon-counting mode spectra from the updated GRB X-ray catalog derived as in \cite{Evans09}.}
 \label{Fig:XraySpecIndex}
 \end{figure*}

\begin{figure} 
 	\hskip -0.6 cm
   \centering
\includegraphics[width=0.5\textwidth]{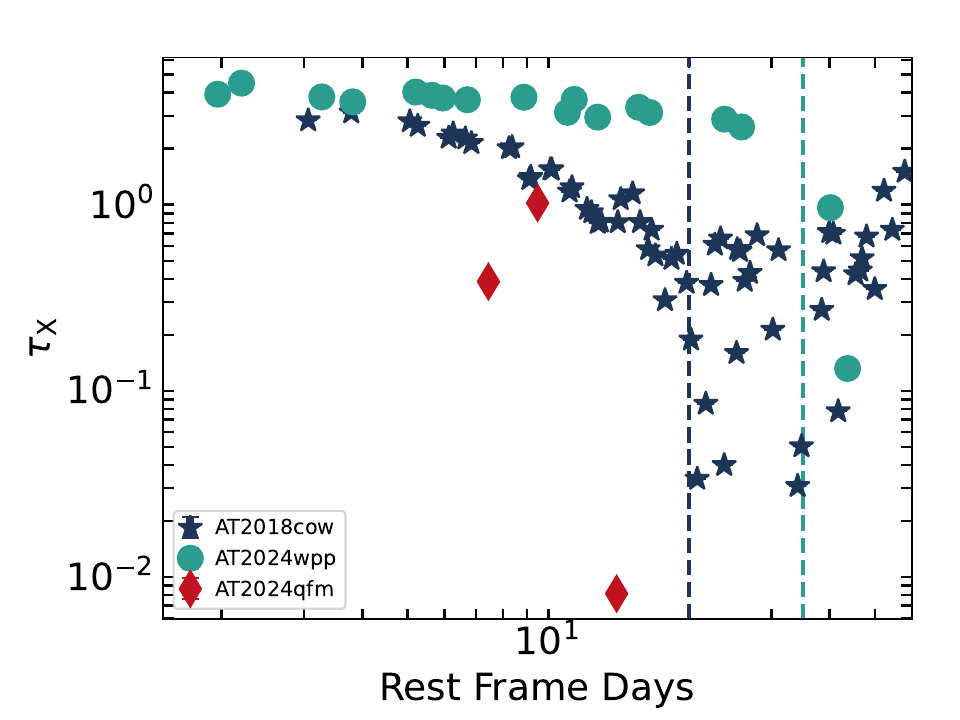}  
 \caption{Evolution of $\tau_{\rm x}$ for AT\,2018cow, AT\,2024wpp, and AT\,2024qfm assuming that the observed UVOIR luminosity ($L_{\rm opt}$) is reprocessed X-ray luminosity ($L_{\rm x}$) from the central source \citep{Metzger22FBOT}. The vertical dashed lines indicate the time of emergence of spectral features in AT\,2018cow and AT\,2024wpp. }
 \label{Fig:tau_x_Metzger22}
 \end{figure}
 
At $\delta t<20$\,days, \at{} is the most X-ray luminous LFBOT  detected so far (Fig. \ref{Fig:XrayFBOTs}), $\approx 20$ times more luminous than AT\,2018cow \citep{Margutti2019-cow} at a similar epoch. While the X-ray and radio emission from LFBOTs appear decoupled based on the broadband SEDs of \S\ref{sec:Xraytoradio-SEDs-FBOTs}, the optical emission (extending from UV to IR, $L_{\rm opt}$) has instead been proposed to be reprocessed high-energy radiation, directly linking it to the X-rays. The reprocessing medium can be situated ``in front'' of a centrally located high-energy source (such as accretion onto a compact object and/or a deeply embedded shock; \citealt{Margutti2019-cow, Metzger22FBOT}), or ``behind'' it (e.g., cooled-down material behind the shock, with the freshly shocked material supplying the high-energy emission; \citealt{Taya2026}). In either scenario, the X-ray and optical emission are fundamentally coupled, and their joint temporal evolution will carry the imprint of the efficiency and/or geometry of the reprocessing medium. 

Figure~\ref{Fig:Lxray-Lbol} presents the evolution of $L_{\rm x}$ and $L_{\rm opt}$ of AT\,2024qfm (left panel). In the right panel, we show the evolution of $L_{\rm x}/L_{\rm opt}$ of \at{} along with that of AT\,2018cow \citep{Margutti2019-cow} and AT\,2024wpp \citep{LeBaron2025,Nayana2025}.  The vertical grey and cyan lines indicate the time of emergence of clear optical spectral features with FWHM of a few $\times 10^{3}\, \rm km\,s^{-1}$ in AT\,2018cow ($\delta t \approx 20$\,days) and AT\,2024wpp ($\delta t \approx 35$\,days), respectively. For both events, the time of $L_{\rm x}/L_{\rm opt} \approx 1$ roughly coincides with the first emergence of these spectral features. A Compton hump/hard X-ray excess was observed in the X-ray spectra of both AT\,2018cow \citep{Margutti2019-cow} and AT\,2024wpp \citep{Nayana2025} at $\delta t_{\rm rest} \approx 8$\,days and   $\delta t_{\rm rest} \approx 50$\,days, respectively. The delayed appearance of Compton hump, $L_{\rm x} \approx L_{\rm opt}$, and spectral features in AT\,2024wpp compared to AT\,2018cow have been attributed to a larger Thomson optical depth ($\tau_{\rm T}$) maintained until later times in AT\,2024wpp relative to AT\,2018cow \citep{LeBaron2025,Nayana2025}, and hence a larger reprocessing efficiency of X-ray radiation into optical maintained until later times in AT\,2024wpp. Consistent with this picture, AT\,2024wpp is also the most luminous UV LFBOT to date.

\at{} exhibits the opposite trend: $L_{\rm x} \approx L_{\rm opt}$ is attained at $\delta t \approx 7$\,days, much earlier than that of both AT\,2018cow ($\delta t \approx 20$\,days) and AT\,2024wpp ($\delta t \approx 35$\,days). Notably, $L_{\rm x}/L_{\rm opt}$ is also high ($\approx$ 8.4 at $\delta t \approx 14$\,days) for \at{} compared to AT\,2018cow and AT\,2024wpp, where the ratios approach to $\approx 1$ at late times. In the next two sections, we compare and investigate $L_{\rm x}$ and $L_{\rm opt}$ evolution across \at{}, AT\,2018cow, and AT\,2024wpp to assess reprocessing efficiency in two scenarios: (i) reprocessing medium in front of the central high energy source \citep[central engine model;][]{Metzger22FBOT} (ii) reprocessing layer behind the shock that produces high-energy emission \citep[shock interaction model;][]{Taya2026}.

\subsubsection{External reprocessor}
\label{SubSec:Metzgermodel-Lx-Lbol-discussion}
\cite{Metzger22FBOT} proposed compact object-star mergers as progenitors for LFBOTs, where the accretion disk launches powerful outflows. Conceptually similar models were proposed by \cite{Tsuna24} and \cite{Klencki2026}. In the \cite{Metzger22FBOT} model, the UVOIR emission is primarily powered by the reprocessing of X-rays from the inner accretion flow onto the compact object. The mass of the disrupted star is distributed between a slowly expanding equatorial outflow and a fast polar outflow, with the latter partially reprocessing the luminous central X-ray source into optical/UV emission. X-rays escaping the polar funnel are observed directly. A key geometric parameter in this framework is $\phi_{\rm 0}$, the fraction of the total solid angle subtended by the slow equatorial ejecta ($\Omega_{\rm slow}=4\pi \phi_0$). The solid angle subtended by the fast-moving ejecta is $\Omega_{\rm fast}=4\pi (1-\phi_{0})$. This parameter controls what fraction of the central $L_{\rm x}$ is intercepted and reprocessed into $L_{\rm opt}$ by the slow outflow, even after the fast polar ejecta has become optically thin. Thus $\phi_{0}$ sets a minimum to $L_{\rm opt}$ or a maximum to $L_{\rm x}/L_{\rm opt}$. The fraction of $L_{\rm x}$ thermalized into $L_{\rm opt}$ by the fast polar ejecta is governed by the optical depth $\tau_{\rm x}$, which decreases with time as the ejecta expands. $\tau_{\rm x}$ is expected to trace the effective optical depth for X-rays. In this framework, the ratio $L_{\rm x}/L_{\rm opt}$ directly encodes the optical depth of the reprocessing medium as $e^{\tau_{\rm x}} \propto \left( \frac{L_{\rm x}+ L_{\rm opt}}{L_{\rm x}} \right)(1-\phi_{0})$. When $\tau_{\rm x} \gg 1$ the X-rays are efficiently thermalized and $L_{\rm x} \ll L_{\rm opt}$, whereas as $\tau_{\rm x} \rightarrow 1$ an increasing fraction of X-rays escape and $L_{\rm x} \sim L_{\rm opt}$.

Following the parametrization presented in \citet[][see their \S 3.2]{Metzger22FBOT}, we plot the effective optical depth to X-ray for \at{} and overplot the same values for AT\,2018cow and AT\,2024wpp \citep{Nayana2025} in Fig \ref{Fig:tau_x_Metzger22}, using isotropic-equivalent values of $L_{\rm x}$ and $L_{\rm opt}$. Following \cite{Metzger22FBOT} and \cite{LeBaron2025}, we use a fiducial value of $\phi_{0}=0.5$ for AT\,2018cow and AT\,2024wpp for illustrative purpose. However, for the observed values of $L_{\rm x}$ and $L_{\rm opt}$ in \at{}, we find no physically plausible solution for $\tau_{\rm T}$ when $\phi_{\rm 0}=0.5$. Reproducing the observed luminosities requires $\phi_{\rm 0} \lesssim 0.1$. Under this constraint, $\tau_{\rm T}$ drops below unity at $\delta t \approx 9$\,days. 

In a homologously expanding ejecta, the evolution of $\tau_{\rm T}$ is expected to be $\tau_{\rm T} \propto t^{-2}$, if effects such as changing ionization states and/or continued deposition of material via winds or other mechanisms are not important. AT\,2024wpp showed roughly constant $\tau_{\rm x}$ until later time with a sudden drop at $\delta t \approx 40$\,days whereas AT\,2018cow followed $\tau_{\rm T} \propto t^{-2}$ evolution \citep{Nayana2025}
. These differences can be due to time-variable $\phi_{0}$ or ejecta ionization levels. Nevertheless, our analysis indicates that \at{} has a reprocessing medium which became optically thin much earlier than AT\,2018cow and AT\,2024wpp. 

\subsubsection{Internal reprocessor}
\label{Subsunsection:LxLopt-Taya model}
CSM interaction models are proposed to explain LFBOTs where the interaction between a low-mass ejecta with a dense CSM can efficiently convert kinetic energy into radiation, powering fast and luminous emission \citep{Pellegrino2022}. A more comprehensive framework of shock-CSM interaction to explain observational signatures of AT\,2018cow is presented in \cite{Taya2026}. In this model, the X-ray emission originates from the freshly shocked electrons, whereas the UVOIR emission results from the X-ray emission reprocessed by the cool downstream material that crossed the shock earlier. X-ray photons escape to the observer with probability min(1, 1/$\tau_{\rm T}$), giving $L_{x}/L_{\rm opt} \approx$ $\tau_{\rm T}$ when the shock is Thomson-thick ($\tau_{\rm T} >$  1). Once $\tau_{\rm T}$ drops below unity, roughly half the X-ray luminosity escapes directly and half is reprocessed giving $L_{\rm x}/L_{\rm opt} \approx 1$. The same optical depth also governs the hard X-ray ($>10$ keV) behavior: while $\tau_{\rm T} \gg 1$, hard X-rays scatter through the downstream largely unattenuated and produce a distinct ``hump" relative to the suppressed soft X-ray continuum, and this hump only disappears once $\tau_{\rm T}$ approaches unity. The time evolution of $\tau_{\rm T}$ itself follows a power law ($\tau_{\rm T} \propto t^{-b}$), where the exponent depends on the CSM density profile and the shock deceleration index \citep[Eqn. 5 of][]{Taya2026}. $\tau_{\rm T}$ and its evolution (their Eqns. 4 and 31) controls both the escape fraction and the persistence of the hard X-ray component.

The transient Compton hump/hard X-ray excess above 10 keV observed in AT\,2018cow at $\delta t \approx 15$\,days naturally fits in the \cite{Taya2026} model. We do not see any clear Compton hump signature in \at{} throughout the X-ray monitoring at $\delta t_{\rm rest} \approx 7-109$\,days although we do see spectral evolution around the time of the re-brightening ($\delta t_{\rm rest} \approx 10$\,days) followed by strong spectral softening. We calculate the evolution of $\tau_{\rm T}$ \citep[using equation 31 from][]{Taya2026} for a CSM density profile of $s=3$ ($\rho_{\rm CSM} \propto r^{-s}$) and shock velocity $\approx 0.2$c. We assume a shock break out time $t_{\rm bo} \approx 2$\,days and the time of disappearance of Compton hump to be $\delta t_{\rm rest} \approx 7.4$\,days as $L_{\rm x} \approx L_{\rm opt}$ by this time. This results in $\tau_{\rm T}$ evolution such that $\tau_{\rm T} \lesssim 1$ at $\delta t_{\rm rest} \approx 10$\,days. Thus, the early occurrence of $L_{\rm x} \approx L_{\rm opt}$ observed in \at{} can be explained in \cite{Taya2026} model for a reasonable set of parameters.

\subsubsection{Emergence of optical spectral features}
\label{Sec:optical-spevctral-features-discussion}
Clear optical spectral features of a few$\times$ 1000 $\rm km\,s^{-1}$ were observed in AT\,2018cow at $\delta t_{\rm rest} \approx 20$\, days while AT\,2024wpp showed features at $\delta t_{\rm rest} \approx 35$\,days. In the shock interaction model \citep{Taya2026}, emission lines are expected to form in the upstream when the temperature is low enough and $\tau_{\rm T} \ll 1$ so that the lines are not broadened via electron scattering. Similarly, in the models invoking a central engine \citep{Metzger22FBOT}, the appearance of emission lines is attributed to the decrease in effective opacity of the reprocessing layer revealing the inner region \citep{LeBaron2025,Aspegren2026}. The delayed appearance of Compton hump in AT\,2024wpp ($\delta t \approx 50$\,days) compared to AT\,2018cow ($\delta t \approx 8$\,days)  is also in line with these observational signatures, indicating large $\tau_{\rm T}$ maintained until late-time in AT\,2024wpp. No spectral features are detected in \at{} from our observations spanning $\delta t \approx 10-127$\,days or the ones presented in \cite{Sevilla2026} at $\delta t \approx 10-41$\,days and in \cite{Fulton2024-qfm-rapid-evolution} at $\delta t \approx 8.6$ and 10.6\,days. However, we notice that the limited signal-to-noise ratio of these optical spectra would prevent the detection of spectral features similar to those of AT\,2018cow \citep{Perley2019,Margutti2019-cow} or AT\,2024wpp \citep{LeBaron2025}.

To summarize the points discussed in \S \ref{SubSec:Metzgermodel-Lx-Lbol-discussion}, \ref{Subsunsection:LxLopt-Taya model}, and \ref{Sec:optical-spevctral-features-discussion}, the observed $L_{\rm x}$/$L_{\rm opt}$ is model-agnostic and can be treated as a proxy for the efficiency of reprocessing where the agent of reprocessing changes in different models. The early occurrence of $L_{\rm x} \approx L_{\rm opt}$ in \at{} could be interpreted as the system reaching the state of low opacity of the reprocessing layer at much earlier times than in AT\,2018cow and AT\,2024wpp. 

\begin{figure*} 
 	\centering	\includegraphics[width=0.46\textwidth]{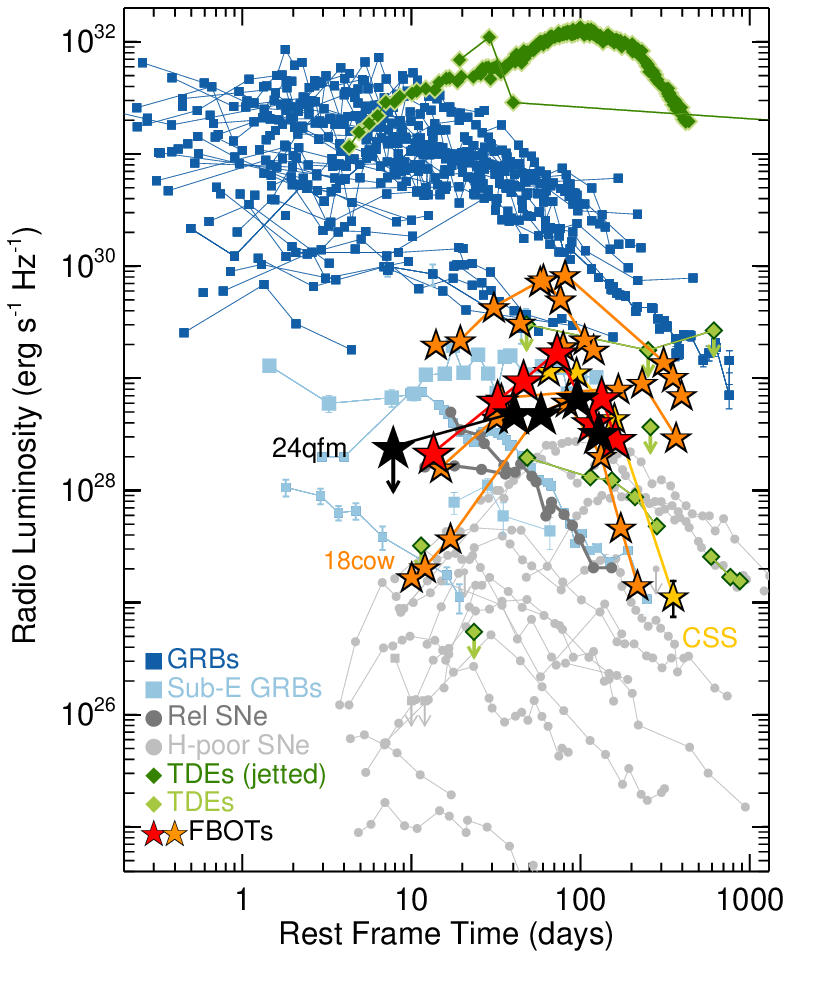} 
    \includegraphics[width=0.47\textwidth]{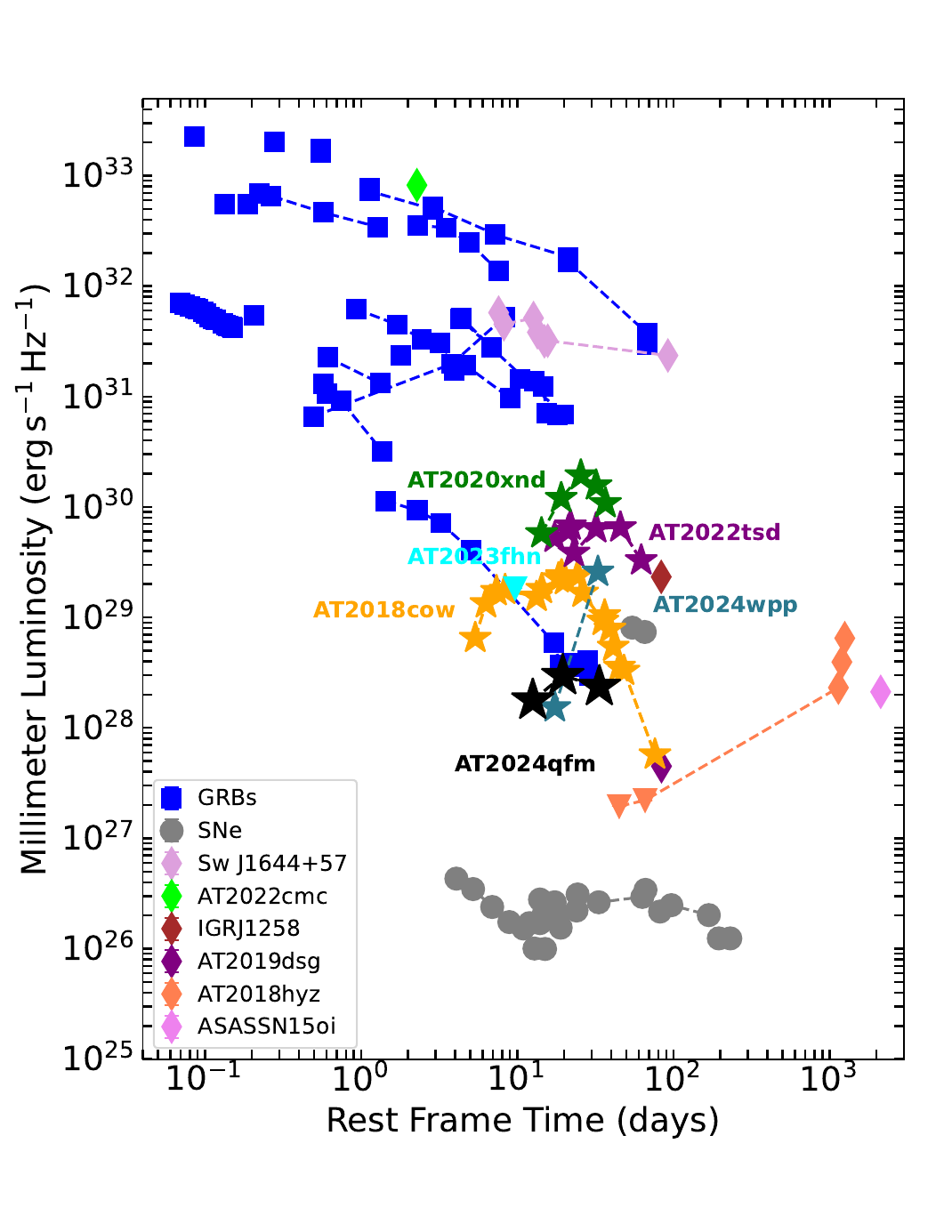} 
  \caption{\textit{Left panel:} Radio light curve of AT\,2024qfm at X-band (8--12 GHz) in the context of other LFBOTs (star symbols) and energetic transients including GRBs, Supernovae, and TDEs \citep[][ and references therein]{Margutti2019-cow, Nayana2025}. All LFBOTs peak at similar times ($\delta t_{\rm rest} \approx 50-100$\,days) in the X-band, with peak luminosities spanning only a factor of $\approx 10$. {\textit{Right panel:}} The mm-band (80--200 GHz) light curves of AT\,2024qfm along with other mm-bright LFBOTs (star symbols); AT\,2018cow \citep{Ho2019-cow}, AT2020xnd \citep{Ho2022-xnd}, AT\,2022tsd \citep{Ho2023-at2022tsd}, AT\,2023fhn \citep{Ho2023-fhn-NOEMA-limit}, and AT\,2024wpp \citep{Nayana2025}. All mm-bright LFBOTs peak at $\delta t_{\rm rest} \approx 1$ month, with a larger dispersion of $\approx 100$ in peak luminosity. Other extragalactic mm-bright transients are also shown for comparison \citep[][and references therein]{Eftekhari2022}.   }
 \label{Fig:radio-xband-mm-lc-comparison}
 \end{figure*}

\subsection{X-ray properties}
\label{Subsec:X-ray-spectra-GRBs v/s LFBOTs}
While the LFBOT X-ray luminosity is similar to that of long GRBs at the same epoch, their spectra are clearly distinct.
In Fig. \ref{Fig:XraySpecIndex} we present the temporal evolution of the 0.3--10 keV spectral index $\beta$, where $F_{\nu}\propto \nu^{-\beta}$, derived from  a uniform X-ray analysis of all LFBOTs observed so far. LFBOTs consistently show harder X-ray spectra than GRB afterglows:  LFBOTs have a typical $\beta \approx 0.5$, while GRB X-ray afterglows are characterized by $\beta\approx 1$. This simple observation has profound implications regarding the nature and physical conditions behind the X-ray emission in LFBOTs, and it is consistent with the inference that the emitting region of the radio-emitting electrons and the X-ray emitting electrons are distinct in LFBOTs (but are typically the same in GRBs at these late epochs).

In the conventional GRB afterglow picture, $\beta = (p-1)/2$ below the cooling frequency, and $\beta = p/2$ above. The softer spectra of GRB afterglows ($\beta \approx 1$) are consistent with $p \approx 2.2$–$2.5$, typical of diffusive shock acceleration \citep{GranotSari2002,Curran2010}. The harder LFBOT spectra would naively require either a hard electron distribution ($p \approx 2$), perhaps driven by magnetic reconnection rather than shock acceleration \citep{Sironi2014}; or the absence of significant synchrotron cooling, implying either a weak magnetic field or continuous re-energization of the emitting electrons. Another alternative is that the X-ray emission in LFBOTs arise from a fast-cooling shock propagating through dense CSM. In this scenario, the X-ray spectrum naturally takes the form $L_\nu \propto \nu^{-1/2}$ (i.e., $\beta = 0.5$) as a result of the superposition of free-free emission from post-shock electrons cooling across a range of temperatures: this is an integrated cooling spectrum rather than a single-temperature plasma \citep[see eqn. B5 from][]{Taya2026}. 

\subsection{Radio and Millimeter properties} 
\label{subsec:radio-comparison}
We place AT\,2024qfm in the phase space of radio luminosity light curves of all LFBOTs and other transients in the cm and mm bands (see Fig. \ref{Fig:radio-xband-mm-lc-comparison}). The peak X-band (8--12\,GHz) spectral luminosity of \at{} is $L \approx 10^{29}\, \rm erg\,s^{-1}\,Hz^{-1}$ at $\delta t_{\rm rest} \approx 96$\,days, falling well in the range of cm band luminosities of LFBOTs. The peak spectral luminosity in the mm-bands is $L_{\rm 94\, GHz} \approx 2.4 \times 10^{29}\, \rm erg\,s^{-1}\,Hz^{-1}$ at $\delta t_{\rm rest} \approx 16$\,days. There are only five other LFBOTs detected in the mm-bands: AT\,2018cow \citep{Ho2019-cow}, AT\,2020xnd \citep{Ho2022-xnd}, 
AT\,2022tsd \citep{Ho2023-at2022tsd}, AT2023vth \citep{Sevilla2026}, and AT\,2024wpp \citep{Nayana2025,Perley2026-wpp}. \at{} marks the least luminous among them. The peak spectral luminosities of LFBOTs in the cm bands cluster around a few times $10^{29}\, \rm erg\,s^{-1}\,Hz^{-1}$ at $\delta t_{\rm rest} \approx 100$\,days whereas the peak luminosity in the mm bands span two orders of magnitude at $t_{\rm rest} \approx 20-40$\,days. This indicates fairly uniform CSM densities at larger radii and more diverse density environments in the immediate vicinity of the progenitor. The inferred CSM densities (see Fig \ref{Fig:density-evolution}) further support this interpretation \citep{Nayana2025,Sevilla2026}. The compilation of SSA peak frequencies for radio-observed FBOTs (Fig \ref{Fig:nup-all-LFBOTs}) reveals a broadly consistent temporal evolution across events, in contrast to CCSNe \citep[e.g.,][]{Bietenholz2021}. This similarity also supports the emerging picture of a common CSM density structure in LFBOTs, suggestive of broadly similar pre-explosion mass-loss histories \citep{Klencki2026}.  

\section{Summary and conclusions} 
\label{Sec:summary}
We present detailed multi-wavelength observations of the LFBOT \at{}. Major findings from our analysis are the following:
\begin{enumerate}
\item The UV-optical emission from \at{} is consistent with a blackbody spectrum with temperatures $T \approx 16,500, 14,200,$ and 10,900 K at $\delta t_{\rm rest} \approx 7.4, 9.4$ and 14.2\,days, respectively. The corresponding blackbody radii are $R_{\rm BB} \approx (8-9) \times 10^{14}$\,cm implying average expansion velocities of $R_{\rm BB}/t \approx (0.02-0.05)c$. Optical spectra obtained at $\delta t \approx 10-127$\,days do not show any transient spectral features. 
    \item AT\,2024qfm shows luminous X-ray emission ($L_{\rm x} \approx 10^{44}\, \rm erg\,s^{-1}$) in the first $\approx 6$\,days, marking it as the most luminous X-ray bright LFBOT at early times. X-ray light curve shows a rapid decline ($L_{\rm x} \propto t^{-6}$) in the first 6 days followed by a re-brightening at $\delta t \approx 15$\,days. X-ray emission also exhibits variability on days timescale.
    \item The X-ray spectra are relatively soft at early times ($F_{\nu} \propto \nu^{-1}$) and get harder around the time of re-brightening with $F_{\nu} \propto \nu^{-0.5}$, a similar pattern to that seen in AT\,2024wpp \citep{Nayana2025}.
    \item Luminous radio emission is detected from AT\,2024qfm with a spectral luminosity of $L_{\rm 11\, GHz} \approx 1.3 \times 10^{29}\, \rm erg\,s^{-1}\,Hz^{-1}$, peaking at $\delta t \approx 96$\,days. Radio emission is broadly consistent with synchrotron radiation arising due to the interaction of a fast-moving ($\Gamma \beta c \approx 0.2$c) shock in a medium of density $n \approx 6 \times 10^{4}\, \rm cm^{-3}$ at $R \approx 10^{16}$\,cm. 
    \item The peaks of the SSA radio spectra evolve from $\nu_{\rm pk} \approx 65$\,GHz to 11 GHz from $\delta t \approx 25$ to 96\,days. However, the SEDs do not evolve significantly between $\delta t \approx 40-58$\,days, indicating multiple outflow components contributing to the overall synchrotron emission.
    \item AT\,2024qfm has a peak mm-band luminosity of $L_{\rm 94\, GHz} \approx 2.4 \times 10^{29}$\,erg\,s$^{-1}$\,Hz$^{-1}$, making it the least luminous LFBOT in the mm bands observed to date. As a population, LFBOTs peak at similar times in the cm band ($\delta t_{\rm rest} \approx 50-100$\,days), with peak luminosities spanning only a factor of $\approx 10$, whereas in the mm bands they all appear to peak at $\delta t_{\rm rest} \approx 1$\,month, with a larger dispersion of $\approx 100$ in peak luminosity. This suggests that LFBOTs share fairly uniform CSM densities at larger radii, but span a much wider range of densities in the immediate vicinity of the progenitor.
    \item A compilation of X-ray spectral properties of all LFBOTs and GRBs indicates that the LFBOTs consistently show harder spectral indexes ($F_{\nu} \propto \nu^{-0.5}$) compared to those of GRBs ($F_{\nu} \propto \nu^{-1}$) even though they exhibit similar luminosities. The harder spectra indicate either the absence of synchrotron cooling or an intrinsically harder electron distribution ($p=2$). Alternatively, free-free emission from a fast-cooling shock could also produce $F_{\nu} \propto \nu^{-0.5}$. 
     \item The X-ray to UV-optical luminosity ratio of AT\,2024qfm is $\gtrsim 1$ at $\delta t \approx 7$\,days. In AT\,2018cow and AT\,2024wpp, the $L_{\rm x}/L_{\rm UVOIR}$ became unity at $\delta t_{\rm rest} \approx 20$ and 35\,days, respectively. In a scenario where UVOIR emission is reprocessed X-ray emission, the reprocessing is least efficient in AT\,2024qfm. These events thus suggest a broad range of line-of-sight reprocessing efficiencies among LFBOTs, which can be tested with contemporaneous X-ray and UV-optical observations and joint modeling of future events.
\end{enumerate}

To summarize, the multiwavelength emission from \at{} falls in the range of observational properties of other X-ray and radio bright LFBOTs. A decade after the discovery of first LFBOT (CSS161010) in 2016, the number of well-monitored LFBOTs still remains small with \at{} being only the twelfth event. In the era of Rubin-LSST (deep drilling fields), ULTRASAT, and UVEX, early time discovery, classification, and rapid multiwavelength follow-up of more LFBOTs will be possible. Building a statistical sample is critically important to understand their intrinsic properties and physical origin as a class. 
\vspace{5mm}

\facilities{Swift(XRT and UVOT), Keck, Palomar, SOAR, Gemini, CXO, NICER, NuSTAR, SMA, NOEMA, JVLA, AMI-LA}
\software{astropy \citep{2013A&A...558A..33A,2018AJ....156..123A},  
          }
\section*{Acknowledgments}
This work was initiated in part at the Aspen Center for
Physics, which is supported by National Science Foundation
grant PHY-2210452. The authors also acknowledge the hospitality of the Kavli Institute for Theoretical Physics at UC Santa Barbara, where portions of this work were completed. This research was supported in part by the National Science Foundation under PHY-2309135.

The Submillimeter Array is a joint project between the Smithsonian Astrophysical Observatory and the Academia Sinica Institute of Astronomy and Astrophysics and is funded by the Smithsonian Institution and the Academia Sinica. We recognize that Maunakea, where the SMA is sited, is a culturally important site for the indigenous Hawaiian people; we are privileged to study the cosmos from its summit.
This research has made use of the NuSTAR Data Analysis Software (NuSTARDAS) jointly developed by the ASI Space Science Data Center (SSDC, Italy) and the California Institute of Technology (Caltech, USA). 
This research has made use of data obtained 
from the Chandra Data Archive provided by the Chandra X-ray Center (CXC).
This research has made use of the XRT Data Analysis Software (XRTDAS) developed under the responsibility
of the ASI Science Data Center (ASDC), Italy.
This work made use of data supplied by the UK Swift Science Data Centre at the University of Leicester.
This research has made use of data and software provided by the High Energy Astrophysics Science Archive Research Center (HEASARC), which is a service of the Astrophysics Science Division at NASA/GSFC.
Some of the data presented herein were obtained at Keck Observatory, which is a private 501(c)3 nonprofit organization operated as a scientific partnership among the California Institute of Technology, the University of California, and the National Aeronautics and Space Administration. The Observatory was made possible by the generous financial support of the W. M. Keck Foundation. 
The authors wish to recognize and acknowledge the very significant cultural role and reverence that the summit of Maunakea has always had within the Native Hawaiian community. We are most fortunate to have the opportunity to conduct observations from this mountain.
Based on observations obtained at the Southern Astrophysical Research (SOAR) telescope, which is a joint project of the Minist\'{e}rio da Ci\^{e}ncia, Tecnologia e Inova\c{c}\~{o}es (MCTI/LNA) do Brasil, the US National Science Foundation’s NOIRLab, the University of North Carolina at Chapel Hill (UNC), and Michigan State University (MSU).
\par R.M. acknowledges support by the National Science
Foundation under award No. AST-2224255, and by NASA under grants 80NSSC22K1587, 80NSSC25K7591 and 80NSSC22K0898. 
J.F.S acknowledges support from NASA Grant NAS8-03060.
B.D.M. acknowledges support from NASA AAG (grant number 80NSSC22K0807), the Fermi Guest Investigator Program (grant number 80NSSC24K0408) and the Simons Foundation (grant number 727700). The Flatiron Institute is supported by the Simons Foundation.
N.L. thanks the LSST-DA Data Science Fellowship Program, which is funded by LSST-DA, the Brinson Foundation, the WoodNext Foundation, and the Research Corporation for Science Advancement Foundation; her participation in the program has benefited this work. 
A.Y.Q.H. acknowledges support from a Sloan Research Fellowship (Award Number FG-2024-21320) from the Alfred P. Sloan Foundation, a Packard Fellowship from the David and Lucile Packard Foundation (Grant Number 2024-77386), National Aeronautics and Space Administration (NASA) grant 80NSSC24K0377, an LSST Scialog Early Science grant from the Research Corporation for Science Advancement, and {\it Hubble Space Telescope} ({\it HST}) grant HST-GO-17477.006-A.
The work of DS was carried out at the Jet Propulsion Laboratory, California Institute of Technology, under a contract with the National Aeronautics and Space Administration (80NM0018D0004).
A.H. is grateful for the support by the Israel Science Foundation (ISF grant 1679/23) and by the United States-Israel Binational Science Foundation (BSF grant 2020203). 
G.M. acknowledges financial support from INAF GO grant 1.05.24.02.03 Bando Ricerca Fondamentale INAF 2024.
H.S. acknowledges partial salary support from a Moore Foundation Postdoctoral Fellowship Grant to Rutgers University. 
K.S.M acknowledges support from NASA Grant 12-EUCLID11-0004, through an appointment to the NASA Postdoctoral Program administered by Oak Ridge Associated Universities.
\appendix
\section{Synchrotron self-absorption model}\label{Appendix-SSA}
We derive the shock radius ($R$) and magnetic field ($B$) for a synchrotron self-absorbed (SSA) spectrum for $\nu_{\rm c} < \nu_{\rm a}$ as a function of peak flux density ($F_{\rm pk}$) and synchrotron self-absorption frequency ($\nu_{\rm a} \equiv \nu_{\rm pk}$). \cite{Ho2022-xnd} derive the expressions for source properties in this regime without the covering factor. We re-derive these equations including explicit dependence on source covering factor. Following the \cite{chevalier1998} model, we assume a shock wave propagating into a CSM of density profile $\rho_{\rm CSM} \propto r^{-s}$. Electrons are accelerated at the shock front in amplified magnetic fields forming a power-law distribution in energy ($E$): $dN = N_{0} E^{-p}dE$ for $E > E_{l}$, where $N$ is the number density of relativistic electrons and $E_{l}=\gamma_{\rm m}m_{\rm e}c^{2}$ is the lowest energy of electrons in the power-law distribution with a  minimum Lorentz factor of $\gamma_{\rm m}$. The SSA frequency ($\nu_{\rm a} \equiv \nu_{\rm pk}$) is defined as the frequency at which asymptotically the optically thick ($F_{\nu}^{\rm thick}$) and thin flux ($F_{\nu}^{\rm thin}$) densities meet. For $\nu_{\rm a} < \nu_{\rm c}$, $F_{\nu}^{\rm thick}$ and $F_{\nu}^{\rm thin}$ are given by \cite{chevalier1998}:
\begin{eqnarray}
\label{eqn:F-thick}
    F_{\nu}^{\rm thick}= \frac{\pi R^{2}}{D^{2}} \frac{c_{5}}{c_{6}} B^{-1/2} \left( \frac{\nu}{2c_{1}} \right) ^{5/2}
    \label{eqn:flux-thick}
\end{eqnarray}

\begin{eqnarray}
\label{eqn:F-thin}
    F_{\nu}^{\rm thin}= \frac{4 \pi f R^{3}}{3 D^{2}} c_{5} N_{0} B^{\frac{p+1}{2}} \left( \frac{\nu}{2c_{1}} \right) ^{\frac{-(p-1)}{2}}
\end{eqnarray}
For spectral regime where $\nu_{\rm c} < \nu < \nu_{\rm a}$, the optically thin flux density would be:
\begin{eqnarray}
\label{eqn:F-thin}
    F_{\nu}^{\rm thin}= \frac{4 \pi f R^{3}}{3 D^{2}} c_{5} N_{0} B^{\frac{p+1}{2}} \left( \frac{\nu_{\rm c}}{2c_{1}}\right)^{\frac{1}{2}}  \left( \frac{\nu}{2c_{1}} \right) ^{\frac{-p}{2}}
    \label{eqn:flux-thin-fast-cooling}
\end{eqnarray}
From equations \ref{eqn:flux-thick} and \ref{eqn:flux-thin-fast-cooling},  $R$ and $B$ in the fast cooling regime are

\begin{eqnarray}
    R = \left( \frac{\pi c_{5}}{c_{6}} \right) ^{\frac{-(p+2)}{2p+7}} \zeta^{\frac{-1}{2p+7}} D^{\frac{2(p+3)}{2p+7}} t^{\frac{1}{2p+7}} f_{\rm eB}^{\frac{-1}{2p+7}} f^{\frac{-1}{2p+7}} F_{\rm pk}^{\frac{p+3}{2p+7}} \nu_{\rm pk}^{\frac{-(2p+5)}{2p+7}}
\end{eqnarray}

\begin{eqnarray}
    B = \left( \frac{\pi c_{5}}{c_{6}} \right)^{\frac{6}{2p+7}} \zeta^{\frac{-4}{2p+7}} D^{\frac{-4}{2p+7}} t^{\frac{4}{2p+7}} f_{\rm eB}^{\frac{-4}{2p+7}} f^{\frac{-4}{2p+7}} F_{\rm pk}^{\frac{-2}{2p+7}} \left( \frac{\nu_{\rm pk}}{2c_{1}} \right)^{\frac{2p+15}{2p+7}}
\end{eqnarray}

The constant $\zeta$ is 
\begin{eqnarray}
    \zeta = \frac{c_{5} (p-2) E_{l}^{p-2}}{6 (2c_{1})^{1/2}} \left(  \frac{18 \pi m_{\rm e} c q}{\sigma_{\rm T}^{2}} \right)^{1/2}
\end{eqnarray}

The constant $c_{1}=6.27 \times 10^{18}$ in cgs units \citep{chevalier1998}. Values of $c_{5}$ and $c_{6}$ are a function of $p$ and are tabulated in \cite{Pacholczyk1970}. 
\begin{eqnarray}
    c_{5} = \frac{\sqrt{3}}{16 \pi} \frac{q^{3}}{m_{e}c^{2}} \frac{p+\frac{7}{3}}{(p+1)} \Gamma \left( \frac{3p-1}{12} \right) \Gamma \left( \frac{3p+7}{12} \right)
    \label{eqn:c5}
\end{eqnarray}
\begin{eqnarray}
    c_{6} = \sqrt{3} \frac{\pi}{72} q m_{e}^{5} c^{10} \left( p + \frac{10}{3} \right) \Gamma \left( \frac{3p+2}{12} \right) \Gamma \left( \frac{3p+10}{12} \right)
\end{eqnarray}
For $p=3$, $c_{5} = 7.51952 \times 10^{-24}$ and $c_{6} = 7.95999\times 10^{-41}$ (in cgs units), and $R$ and $B$ can be written as

\begin{eqnarray}
    R = (7.7570 \times 10^{15}) \, f_{\rm eB}^{\frac{-1}{13}} \left( \frac{f}{0.5}\right)^{\frac{-1}{13}}  \left(  \frac{F_{\rm pk}}{\rm Jy}\right)^{\frac{6}{13}} \left( \frac{D}{\rm Mpc} \right)^{\frac{12}{13}} \left( \frac{\nu_{\rm pk}}{5\, \rm GHz} \right)^{\frac{-11}{13}} \left( \frac{t}{100\, \rm days} \right)^{\frac{1}{13}} \, \rm cm    
\end{eqnarray}

\begin{eqnarray}
    B = (0.3545) \, f_{\rm eB}^{\frac{-4}{13}} \left( \frac{f}{0.5} \right)^{\frac{-4}{13}} \left( \frac{F_{\rm pk}}{\rm Jy} \right)^{\frac{-2}{13}} \left( \frac{D}{\rm Mpc} \right)^{\frac{-4}{13}} \left( \frac{\nu_{\rm pk}}{5\, \rm GHz} \right)^{\frac{21}{13}} \left( \frac{t}{100\, \rm days} \right)^{\frac{4}{13}} \, \rm G
\end{eqnarray}


\section{UV/Optical Photometry}\label{AppendixUVOT-photometry}

\restartappendixnumbering
\setcounter{table}{0}
\startlongtable
\begin{deluxetable*}{lcccccc}
\tablecaption{UV/Optical photometry of AT\,2024qfm (AB mag).}
\tablehead{
\colhead{Facility} & \colhead{Start date} & \colhead{Time$^{\rm{a}}$} & \colhead{Filter} & \colhead{mag$^{\rm{b}}$}  \\
  & (MJD) & (d) &    }
\startdata
\textit{Swift}-UVOT   & 60523.40 & 8.99  & $w1$ & $19.90 \pm 0.30$ \\
\textit{Swift}-UVOT   & 60523.49 & 9.08  & $w2$ & $21.05 \pm 0.09$ \\
\textit{Swift}-UVOT   & 60523.43 & 9.02  & $m2$ & $20.76 \pm 0.11$ \\
\textit{Swift}-UVOT   & 60525.99 & 11.58 & $w1$ & $22.05 \pm 0.22$ \\
\textit{Swift}-UVOT   & 60525.99 & 11.58 & $w2$ & $21.68 \pm 0.09$ \\
\textit{Swift}-UVOT   & 60525.99 & 11.58 & $m2$ & $21.65 \pm 0.13$ \\
\textit{Swift}-UVOT   & 60530.55 & 16.14 & $w1$ & $23.43 \pm 0.36$ \\
\textit{Swift}-UVOT   & 60531.86 & 17.45 & $w2$ & $23.06 \pm 0.33$ \\
\textit{Swift}-UVOT   & 60531.86 & 17.45 & $m2$ & $23.06 \pm 0.38$ \\
\enddata
\tablecomments{$^{\rm{a}}$ With respect to first light in the observer frame. $^{\rm{b}}$ Host subtracted and extinction corrected.
\label{Tab:UVOT-photometry}}
\end{deluxetable*}

\section{X-ray observations logs and tables}\label{AppendixXray}

\startlongtable
\begin{deluxetable*}{lcccccc}
\tablecaption{X-ray observations of AT\,2024qfm with \nustar\, and \chandra\, (PI Margutti).}
\tablehead{
\colhead{Instrument} & \colhead{Start date} & \colhead{Mid time$^{\rm{a}}$} & \colhead{Obs ID} & \colhead{Exposure Time$^{\rm{b}}$} & \colhead{Net Count-rate$^{\rm{c}}$} & \colhead{Detection}  \\
  & (yyyy/mm/dd) & (d) &  & (ks)  & ($10^{-4}\,\rm{c\,s^{-1})}$ &
  \colhead{significance ($\sigma$)$^{\rm{c}}$}  }
\startdata
CXO/ACIS-S &   2024-08-07  & 14.75  & 28066  &  19.8 &  $94.9\pm 6.9$ & 79.8\\
CXO/ACIS-S & 2024-08-15& 23.63 & 28067  &19.8 & $34.4\pm 4.2$ & 29.7\\
CXO/ACIS-S & 2024-08-21 & 28.92 & 28068 & 19.8 & $16.3\pm 2.9 $ & 14.9 \\
CXO/ACIS-S &  2024-09-02 & 40.86  & 28069 & 20.8 & $9.47\pm 2.15 $ & 9.7\\
CXO/ACIS-S & 2024-09-30 & \rdelim\}{2}{*}[\, 68.83 ] & 28070 & 22.8 &  \rdelim\}{2}{*}[\,$<1.00$ ]  & -- \\
CXO/ACIS-S & 2024-09-30 &  & 30474 & 16.9 & & -- \\
CXO/ACIS-S & 2024-10-22 & \rdelim\}{2}{*}[\, 80.04 ]& 30584 &17.8  &  \rdelim\}{2}{*}[\,$<1.04$ ]  & -- \\
CXO/ACIS-S & 2024-11-05 & & 28071 & 20.6& & -- \\
NuSTAR & 2024-08-02 &  10.19 & 81002403002 &  43.8/43.5 & $59.8\pm6.8$/$42.5\pm6.2$  & 8.8/6.8 \\
NuSTAR & 2024-08-05 & 14.15 & 81002403004 &    57.4/57.4 & $16.4\pm 3.0$/$15.2\pm3.3$ & 5.5/4.6 \\
NuSTAR & 2024-08-12 &  21.06 & 81002403006 &  51.7/51.3  &   $4.84\pm 3.07$ / $10.4\pm3.1$  & 1.6/2.9\\
\enddata
\tablecomments{$^{\rm{a}}$ With respect to first light in the observer frame. $^{\rm{b}}$ For NuSTAR we report the exposures for the A and B modules, respectively, after removing the interval of times severely affected by solar flares. $^{\rm{c}}$ 0.5-8 keV for \chandra; 3-20 keV for \nustar\, for the A/B modules, respectively, for a source region of $50''$ radius.
\label{Tab:Xraylog}}
\end{deluxetable*}

\startlongtable
\begin{deluxetable*}{lcccccc}
\tablecaption{Unabsorbed X-ray fluxes of AT\,2024qfm at multiple epochs.}
\tablehead{
\colhead{Instrument} & \colhead{Start date} & \colhead{Mid time$^{\rm{a}}$} & \colhead{Unabsorbed Flux}   \\
  & (MJD) & (d) &   ($10^{-14}$ erg cm$^{-2}$ s$^{-1}$)   }
\startdata
XRT   & 60523.36 & 8.96   & $65.09 \pm 15.41$ \\
XRT   & 60523.42 & 9.02   & $44.56 \pm 12.69$ \\
XRT   & 60523.54 & 9.14   & $45.53 \pm 13.72$ \\
XRT   & 60524.55 & 10.15  & $31.81 \pm 7.74$ \\
XRT   & 60532.34 & 17.94  & $14.09 \pm 4.72$ \\
XRT   & 60535.32 & 20.92  & $13.07_{-6.03}^{+8.17}$ \\
XRT   & 60527.56 & 13.16  & $<16.03$ \\
XRT   & 60648.43 & 134.03 & $<1.37$ \\
CXO   & 60529.15 & 14.75  & $31.55_{-2.66}^{+2.33}$ \\
CXO   & 60538.03 & 23.63  & $8.42_{-1.01}^{+1.26}$ \\
CXO   & 60543.32 & 28.92  & $3.87_{-0.64}^{+0.96}$ \\
CXO   & 60555.26 & 40.86  & $2.47_{-0.48}^{+1.19}$ \\
CXO   & 60583.47 & 69.07  & $<0.26$ \\
CXO   & 60612.61 & 98.21  & $<0.27$ \\
NICER & 60527.79 & 13.39  & $27.41_{-8.45}^{+7.10}$ \\
NICER & 60528.48 & 14.08  & $39.61_{-10.92}^{+8.01}$ \\
NICER & 60529.42 & 15.02  & $36.13_{-9.35}^{+9.35}$ \\
NICER & 60530.45 & 16.05  & $18.11_{-4.99}^{+5.76}$ \\
NICER & 60531.58 & 17.18  & $30.05_{-7.25}^{+6.08}$ \\
NICER & 60532.51 & 18.11  & $54.68_{-26.64}^{+12.59}$ \\
EP    & 60526.04 & 11.64  & $6.90_{-2.30}^{+6.30}$ \\
\enddata
\tablecomments{$^{\rm{a}}$ With respect to first light in the observer frame.
\label{Tab:UnabsorbedXrayFlux}}
\end{deluxetable*}

\section{Radio observation logs and data Tables}\label{AppendixRadio}

\startlongtable
\begin{deluxetable*}{ccccccccc}
\tablecaption{SMA and NOEMA observations of AT\,2024qfm (PIs Edo Berger [SMA] and Anna Ho [NOEMA] ). }
\tablehead{
\colhead{Start Date} & \colhead{Project ID}  & \colhead{Centroid MJD} & \colhead{Time$^{\rm{a}}$} & \colhead{Frequency}  & \colhead{Flux Density$^{\rm{b}}$} \\
 (yyyy/mm/dd) &  &  & (d) & (GHz) & ($\mu$Jy) &
 }
\startdata
2024-08-03  & POETS & 60525.00  & 10.59 & 225 & $< 570$ \\
2024-08-05	& D24AA & 60527.00	& 12.59	&	94	&	88$\pm$22 	\\
2024-08-12	& D24AA & 60534.00	& 19.59	&	94	&	147$\pm$33		\\
2024-08-26	& D24AA & 60548.00	& 33.59	&	94	& 117$\pm$21	\\	
2024-09-20 & D24AA & 60573.00 & 58.59 & 94 & $< 60$ \\
\enddata
\tablecomments{$^{\rm{a}}$ With respect to first light MJD 60514.41. $^{\rm{b}}$ The flux density upper limits are $3\sigma$
\label{Tab:radio-NOEMA}}
\end{deluxetable*}

\startlongtable
\begin{deluxetable*}{ccccccccc}
\tablecaption{JVLA observations of AT\,2024qfm (PI Nayana A.J.) }
\tablehead{
\colhead{Start Date} & \colhead{Project ID}  & \colhead{Centroid MJD} & \colhead{Time$^{\rm{a}}$} & \colhead{Frequency}  & \colhead{Flux Density$^{\rm{b}}$} \\
 (yyyy/mm/dd) &    &  & (d) & (GHz) & ($\mu$Jy) &
 }
\startdata
2024-08-10 & 23B-138	& 60524.00	 &  9.59	&	10.0   &	$<$15		\\
2024-08-10 & 24B-081	& 60532.00	 &  17.59	&	15.0   &	$<$15		\\
2024-08-20 & 24B-081	& 60542.27	 &  27.86	&   17.0   &	36$\pm$5 \\
2024-08-23 & 24B-081	& 60545.25   &  30.84	&	5.00   &	$<$18		\\
2024-09-11 & 24B-081   & 60564.13   &  49.72	&	9.00   &  30$\pm$7 \\
2024-09-11 & 24B-081   & 60564.13   &  49.72	&	11.00  &	39$\pm$4 \\
2024-09-11 & 24B-081   & 60564.13   &  49.72	&	13.00  &	45$\pm$4 \\
2024-09-11 & 24B-081   & 60564.13   &  49.72	&	15.00  &	59$\pm$6 \\
2024-09-11 & 24B-081   & 60564.13   &  49.72	&	17.00  &	75$\pm$8 \\ 
2024-09-14 & 24B-081   & 60567.31   &  52.90	&	31.00  &	77$\pm$8 \\ 
2024-09-14 & 24B-081   & 60567.31   &  52.90	&	35.00  &	57$\pm$9 \\ 
2024-10-03 & 24B-081   & 60586.07   &  71.66	&	07.00  &	24$\pm$5 \\  
2024-10-03 & 24B-081   & 60586.07   &  71.66	&	09.00  &	30$\pm$10 \\  
2024-10-03 & 24B-081   & 60586.07   &  71.66	&	13.00  &	49$\pm$10  \\ 
2024-10-03 & 24B-081   & 60586.07   &  71.66	&	15.00  &	51$\pm$12  \\ 
2024-10-03 & 24B-081   & 60586.07   &  71.66	&	17.00  &	50$\pm$12  \\ 
2024-10-02 & 24B-081   & 60585.07   &  70.66	&	31.00  &	67$\pm$10  \\ 
2024-10-02 & 24B-081   & 60585.07   &  70.66	&	35.00  &	64$\pm$10  \\ 
2024-11-18 & 24B-081   & 60632.02   &  117.61	&	5.00   &	32$\pm$06  \\
2024-11-18 & 24B-081   & 60632.02   &  117.61	&	9.00   &	41$\pm$06  \\
2024-11-18 & 24B-081   & 60632.02   &  117.61	&	13.00  &	36$\pm$05  \\
2024-11-20 & 24B-081   & 60634.09   & 119.68 	&	31.00  &	$<$ 45  \\
2024-11-20 & 24B-081   & 60634.09   & 119.68 	&	35.00  &	$<$ 45  \\
2024-12-28 & 24B-081   & 60671.98   &  157.57	&	10.00  &	20$\pm$5  \\
\enddata
\tablecomments{$^{\rm{a}}$ With respect to first light MJD 60514.41 in observer frame. $^{\rm{b}}$ The uncertainties on flux density measurements are map rms values (1$\sigma$). The flux density upper limits are $3\sigma$.
\label{tab:radio-vla}}
\end{deluxetable*}

\startlongtable
\begin{deluxetable*}{ccccccccc}
\tablecaption{AMI observations of AT\,2024qfm}
\tablehead{
\colhead{Start Date} &  \colhead{Centroid MJD} & \colhead{Time$^{\rm{a}}$} & \colhead{Frequency}  & \colhead{Flux Density$^{\rm{b}}$} \\
 (yyyy/mm/dd) &    & (d) & (GHz) & (mJy) &
 }
\startdata
2024-08-10  & 60532 & 17.59 & 15 & $<0.108$ \\
2024-08-20  & 60542 & 27.59 & 15 & $<0.138$ \\
2024-10-19  & 60602 & 87.59 & 15 & $<0.189$ \\
2024-10-27  & 60610 & 95.59 & 15 & $<0.246$ \\
2024-11-08  & 60622 & 107.59 & 15 & $<0.183$ \\
2024-11-14  & 60628 & 113.59 & 15 & $<0.255$ \\
2024-11-16  & 60630 & 115.59 & 15 & $<0.48$ \\
2025-01-04  & 60679 & 164.59 & 15 & $<0.198$ \\
2025-01-20  & 60695 & 180.59 & 15 & $<0.216$ \\
2025-02-06  & 60712 & 197.59 & 15 & $<0.234$ \\
2025-03-04  & 60738 & 223.59 & 15 & $<0.219$ \\
2025-03-08  & 60742 & 227.59 & 15 & $<0.33$ \\
2025-03-29  & 60763 & 248.59 & 15 & $<0.108$ \\
2025-04-27  & 60792 & 277.59 & 15 & $<0.348$ \\
2025-06-08  & 60834 & 319.59 & 15 & $<0.306$ \\
2025-06-09  & 60835 & 320.59 & 15 & $<0.111$ \\
2025-06-16  & 60842 & 327.59 & 15 & $<0.084$ \\
2025-06-30  & 60856 & 341.59 & 15 & $<0.255$ \\
\enddata
\tablecomments{$^{\rm{a}}$ With respect to first light MJD 60514.41 in observer frame.  $^{\rm{b}}$ The flux density upper limits are $3\sigma$.
\label{tab:radio-AMI}}
\end{deluxetable*}

\startlongtable
\begin{deluxetable*}{ccccccccc}
\tablecaption{Compilation of inferred physical parameters of all LFBOTs from radio SED modeling}
\tablehead{
\colhead{Source$^{\rm{a}}$} & \colhead{Phase$^{\rm{b}}$}  & \colhead{$\nu_{\rm pk}$} & \colhead{$F_{\rm pk}$} & \colhead{R}  & \colhead{$\Gamma \beta$} & B & n & U \\
  & (d)   & (GHz) & (mJy) & ($10^{16}$\,cm) & (c) & (G) & (cm$^{-3}$) & (10$^{48}$\,erg)
 }
\startdata
AT\,2018cow & 22 & 100 & 94 & 0.7 & 0.13 & 6 & 3$\times 10^{5}$ & 4 		\\
& 10.3 & $166.5\pm8.4$ & $31.7\pm1.81$ & $0.30\pm0.02$ & $0.113\pm0.006$ & $11.73\pm$0.59 & $1.1 \times 10^{6}$ & $0.54\pm0.05$ \\
& 13.4 & $150.3\pm8.7$ & $41.64\pm2.92$ & $0.38\pm0.02$ & $0.109\pm0.007$ & $10.29\pm0.60$ & $9.1 \times 10^{5}$ & $0.83\pm0.09$ \\
& 17.4 & $135.8\pm10.4$ & $54.5\pm5.4$ & $0.47\pm0.04$ & $0.105\pm0.009$ & $9.04\pm0.70$ & $7.6 \times 10^{5}$ & $1.27\pm0.18$\\
& 22.0 & $100.0\pm2.4$ & $95.12\pm1.59$ & $0.83\pm0.02$ & $0.145\pm0.004$ & $6.28\pm0.15$ & $1.9 \times 10^{5}$ & $3.39\pm0.10$ \\
& 82.8 & $15.66\pm0.76$ & $11.35\pm0.31$ & $1.84\pm0.90$ & $0.086\pm0.004$ & $1.229\pm0.059$ & $2.1 \times 10^{4}$ & $1.65\pm0.10$ \\
& 91.5 & $12.41\pm0.61$ & $9.47\pm0.26$ & $2.11\pm0.10$ & $0.089\pm0.004$ & $0.99\pm0.05$ & $1.3 \times 10^{4}$ & $1.67\pm0.10$ \\
& 132.6 & $5.40\pm0.32$ & $4.96\pm0.20$ & $3.50\pm0.21$ & $0.102\pm0.006$ & $0.46\pm0.03$ & $2.1 \times 10^{3}$ & $1.75\pm0.13$ \\
& 150.2 & $4.08\pm0.27$ & $3.98\pm0.18$ & $4.15\pm0.28$ & $0.107\pm0.007$ & $0.36\pm0.02$ & $1.2 \times 10^{3}$ & $1.78\pm0.15$ \\
CSS161010   & 	69 & 5.6$\pm$0.2 & 8.8$\pm$0.2 & 9.5$\pm$0.4 & 0.53$\pm$0.02 & 0.38$\pm$0.02 & 47$\pm$5 & 29$\pm$1	\\
& 99 & 4.4$\pm$0.1 & 12.2$\pm$0.3 & 14.1$\pm$0.5 & 0.55$\pm$0.02 & 0.29$\pm$0.01 & 25$\pm$2 & 56$\pm$2 \\
& 357 & 0.63$\pm$0.07 & 1.2$\pm$0.1 & 33$\pm$4 & 0.36$\pm$0.04 & 0.052$\pm$0.006 & 1.9$\pm$0.6 & 24$\pm$4 \\
AT\,2018lug & 81 & $\lesssim 8$ & $\gtrsim 0.364$ & $\gtrsim 8.0$ & $\gtrsim 0.38$ & $\lesssim 0.51$ & 190 & 34 \\
AT\,2020xnd & 38 & 61.7$\pm$1.2 & 1.38$\pm$0.04 & 1.11$\pm$0.03 & 0.138$\pm$0.003 & 2.28 $\pm$ 0.04 & (19.8$\pm$0.4) $\times 10^{3}$ & 28.8$\pm$1.7   \\
& 46 & 52.6$\pm$2.2 & 1.05$\pm$0.04 & 1.14$\pm$0.05 & 0.119$\pm$0.005 & 2.00$\pm$0.08 & (29.7$\pm$3.7)$\times 10^{3}$ & 1.5$\pm$0.2 \\
AT\,2020mrf & 261 & 7.44$^{+0.44}_{-0.52}$ & $0.397^{+0.054}_{-0.021}$ & $\approx$ 5.1 & 0.07$-$0.08 & $\approx$ 0.5 & $\approx 3.5 \times 10^{3}$ & $\approx 17$ \\
 & 417$^{\rm{c}}$ & 4.82$^{+1.36}_{-1.18}$ & $0.101^{+0.008}_{-0.008}$ & $4.1^{+1.4}_{-0.9}$ & $0.04^{+0.01}_{-0.01}$ & $0.37^{+0.11}_{-0.09}$ & $(7.6^{+13.1}_{-5.1})\times10^{3}$ & $4.9^{+1.6}_{-1.1}$ \\
AT\,2022tsd & 40 & $\lesssim 100$ & $\gtrsim 0.3$ & $\gtrsim 0.6$ & $\gtrsim 0.06$ & $\lesssim 6$ & $\lesssim 6 \times 10^{5}$ & $\lesssim 3$ \\
AT\,2022adfc & 32 & $> 19$ & $> 0.052$ & $< 1.4$ & $< 0.21$ & $> 1.49$ & $> 5.2 \times 10^{3}$ & $< 1.7$ \\
AT\,2023hkw & 44 & $> 15$ & $> 0.10$ & $< 4.2$ & $< 0.44$ & $> 0.95$ & $> 0.40 \times 10^{3}$ & $< 16$ \\
& 117 & $> 8$ & $> 0.085$ & $< 9.2$ & $< 0.37$ & $> 0.5$ & $> 0.16\times10^{3}$ & $< 47$ \\  
AT\,2023vth & 81 & 9.1$\pm$0.7 & 2.6$\pm$0.2 & 7.1$\pm$0.6 & 0.31$\pm$0.03 & 0.58$\pm$0.04 & (0.31$\pm$0.09)$\times 10^{3}$ & 31$\pm$3 \\
& 110 & 2.9$\pm$0.2 & 1.50$\pm$0.13 & 17.6$\pm$1.3 & 0.53$\pm$0.04 & 0.19$\pm$0.01 & 10$\pm$3 & 50$\pm$6\\
AT\,2024fhn & 90 & 3.6$\pm$0.2 & $0.21\pm0.02$ &  13$\pm$1 & 0.72$\pm$0.03 & 0.24$\pm$0.02 & 20$\pm$2 & 22$\pm$0.3 \\
& 138 & 5.0$\pm$0.7 & $0.35\pm0.05$ & 12$\pm$1 & 0.42$\pm$0.03 & 0.31$\pm$0.01 & 100$\pm$20 & 29$\pm$5 \\
AT\,2024wpp & 32 & 90$^{+5}_{-5}$ & 2.697$^{+0.077}_{-0.069}$ & 0.56$^{+0.03}_{-0.03}$ & 0.067$^{+0.004}_{-0.004}$ & 4.1$^{+0.4}_{-0.4}$ & 0.31$^{+0.09}_{-0.07} \times 10^{6}$ & 0.76$^{+0.01}_{-0.01}$ \\
& 46 & $21^{+3}_{-2}$ & $1.524^{+0.062}_{-0.052}$ & $2.91^{+0.36}_{-0.36}$ & $0.244^{+0.03}_{-0.03}$ & $1.3^{+0.2}_{-0.2}$ & $2.5^{+1.8}_{-0.0}\times 10^{3}$ & $11.44^{+1.33}_{-1.33}$ \\
& 73 & $8^{+0.5}_{-0.5}$ & $1.673^{+0.074}_{-0.068}$ & $7.88^{+0.65}_{-0.56}$ & $0.418^{+0.035}_{-0.030}$ & $0.5^{+0.03}_{-0.03}$ & $1.23^{+0.38}_{-0.31} \times 10^{2}$ & $33.12^{+3.77}_{-3.10}$ \\
& 118 & $6^{+4}_{-2}$ & $0.624^{+0.159}_{-0.112}$ & $6.04^{+2.87}_{-2.23}$ & $0.198^{+0.094}_{-0.073}$ & $0.4^{+0.3}_{-0.1}$ & $4.2^{+4.6}_{-0.3} \times 10^{2}$ & $11.68^{+7.6}_{-3.9}$ \\
\enddata
\tablecomments{$^{\rm{a}}$ References: AT\,2018cow \citep{Ho2019-cow,Margutti2019-cow}, CSS161010 \citep{coppejans2020}, AT\,2018lug \citep{Ho2020-AT2018lug}, AT\,2020xnd \citep{Ho2022-xnd}. \\ $^{\rm{b}}$ With respect to first light, in the observer frame. \\
$^{\rm{c}}$ Shock parameters at 417\,days are not reported by \cite{Yao2021-mrf}; we calculate them from the SED fit using the same formalism and assumptions as \cite{Yao2021-mrf} adopt at 261\,days.
\label{tab:spectral-paramaters-all-LFBOTs}}
\end{deluxetable*}


\bibliography{AT2024qfm_v1}{}
\bibliographystyle{aasjournal}

\end{document}